\documentclass[prl,twocolumn,nopacs,nofootinbib,superscriptaddress, floatfix]{revtex4-2}

\usepackage{amsmath}
\usepackage{amsfonts}
\usepackage{graphicx}
\usepackage{dcolumn}
\usepackage{bm}
\usepackage{xcolor}

\newcommand{\nocontentsline}[3]{}
\newcommand{\tocless}[2]{\bgroup\let\addcontentsline=\nocontentsline#1{#2}\egroup}

\usepackage{hyperref}
\definecolor{MidnightBlue}{HTML}{191970}
\hypersetup{
	colorlinks=true,
	allcolors=MidnightBlue
}

\begin{document}
\preprint{APS/123-QED}

\title{Schr\"odinger dynamics and Berry phase of undulatory locomotion}

\author{Alexander E. Cohen}
\address{Department of Mathematics, 
Massachusetts Institute of Technology, 
77 Massachusetts Avenue, 
Cambridge, MA 02139}
\address{Department of Chemical Engineering, 
Massachusetts Institute of Technology, 
25 Ames Street, 
Cambridge, MA 02142}
\author{Alasdair D. Hastewell}
\address{Department of Mathematics, 
Massachusetts Institute of Technology,  
77 Massachusetts Avenue, 
Cambridge, MA 02139}

\author{Sreeparna  Pradhan}
\address{Picower Institute for Learning and Memory, Department of Brain and Cognitive Sciences, 
Massachusetts Institute of Technology, 
43 Vassar Street,
Cambridge, MA 02139}

\author{Steven W. Flavell}
\address{Picower Institute for Learning and Memory, Department of Brain and Cognitive Sciences, 
Massachusetts Institute of Technology, 
43 Vassar Street,
Cambridge, MA 02139}

\author{J\"orn Dunkel}
\email{dunkel@mit.edu}
\address{Department of Mathematics, 
Massachusetts Institute of Technology, 
77 Massachusetts Avenue, 
Cambridge, MA 02139}

\date{\today}

\begin{abstract}
Spectral mode representations play an essential role in various areas of physics, from quantum mechanics to fluid turbulence, but they are not yet extensively used to characterize and describe the behavioral dynamics of living systems. Here, we show that mode-based linear models inferred from experimental live-imaging data can provide an accurate low-dimensional description of undulatory locomotion in worms, centipedes, robots, and snakes. By incorporating physical symmetries and known biological constraints into the dynamical model, we find that the shape dynamics are generically governed by Schr\"odinger equations in mode space. The eigenstates of the effective biophysical Hamiltonians and their adiabatic variations enable the efficient classification and differentiation of locomotion behaviors in natural, simulated, and robotic organisms using Grassmann distances and Berry phases. While our analysis focuses on a widely studied class of biophysical locomotion phenomena, the underlying approach generalizes to other physical or living systems that permit a mode representation subject to geometric shape constraints.
\end{abstract}

\maketitle


Undulatory propulsion is the natural locomotion strategy~\cite{gray1953undulatory, cohen2010swimming} of many aquatic and terrestrial animals, from worms~\cite{niebur1991theory, majmudar2012experiments,kudrolli2019burrowing, sznitman2010material, shen2011undulatory} and fish~\cite{smits2019undulatory,thewissen1997locomotor} to lizards~\cite{maladen2011mechanical,chong2022coordinating} and snakes~\cite{guo2008limbless, hu2009mechanics}. 
The mechanical wave patterns that drive undulatory motion reflect an animal's behavioral state~\cite{flavell2020behavioral}, providing a macroscopic physical readout of the underlying biochemical and neuronal excitations. 
Recent advances in automated live-imaging~\cite{hong2015automated,pokala2022recording} enable simultaneous observations of macroscopic locomotion dynamics and microscopic cellular activity~\cite{cermak2020whole,shipley2014simultaneous,leifer2011optogenetic,prevedel2014simultaneous,bozek2021markerless,liu2018temporal}, producing rapidly growing multi-scale data sets~\cite{hebert2021wormpose} that have to be tracked~\cite{mathis2020deep,pereira2019fast,mathis2018deeplabcut} and translated into predictive and interpretable models. 
Despite recent major progress in the experimental characterization~\cite{cermak2020whole,shipley2014simultaneous,leifer2011optogenetic,prevedel2014simultaneous,bozek2021markerless} and biophysical description of specific organisms~\cite{stephens2008dimensionality,costa2019adaptive,ahamed2021capturing,majmudar2012experiments,maladen2011mechanical,hosoi2015beneath,ozkan2021collective,berman2014mapping, lauga2009hydrodynamics, astley2020surprising}, a quantitative model inference framework for comparing experimentally observed undulatory dynamics within and across species has yet to be developed. 
In addition to providing unifying biophysical insights spanning different animal kingdoms, such a framework would also allow for a direct comparison of living systems with computational models~\cite{boyle2012gait,sarma2018openworm} and biomimetic robotic devices~\cite{marvi2014sidewinding,aguilar2016review}.
\par
Here, we use spectral mode representations to identify symmetry-constrained dynamical models that can capture and distinguish the undulatory locomotion of worms (\textit{Caenorhabditis elegans})~\cite{flavell2020behavioral}, neuro-mechanical worm models~\cite{boyle2012gait}, Mojave shovel-nosed snakes (\textit{Chionactis occipitalis})~\cite{schiebel2019mechanical}, mechanical snakes, and centipedes (\textit{Lithobius forficatus}). 
Compared with traditional continuum descriptions of undulatory shape-deformations in position space, formulating locomotion models in mode space~\cite{goldschmidt2021bilinear, schmid2010dynamic, tu2013dynamic,romeo2021learning} offers several theoretical and practical advantages: (\textit{i}) high-dimensional experimental data can be efficiently compressed to obtain an interpretable low-dimensional representation; (\textit{ii}) the mode dynamics reduces to a system of linear ordinary differential equations (ODEs); (\textit{iii})~physical symmetries and biological constraints can be efficiently encoded through the structure of the dynamical matrix; (\textit{iv}) all model parameters can be directly inferred from experimental data using ODE sensitivity methods~\cite{ma2021comparison,rackauckas2020universal} that exploit the imposed matrix structure~\cite{sorber2015structured}. 
In particular, for undulatory locomotion, we find that translational invariance, rotational invariance, and length constraints generically lead to a Schr\"odinger equation~\cite{schrodinger1926undulatory} in mode space.
Similar to the characterization of quantum systems in terms of their spectra and eigenstates~\cite{griffiths2018introduction}, 
the eigenspaces of the effective Hamiltonians enable an efficient classification of the locomotion dynamics of worms, snakes, robots, and computational models.
Furthermore, transitions between animal behavioral states are encoded in the time evolution of the Hamiltonian  and thus can be detected using Berry phases~\cite{berry1984quantal}.
While our discussion focuses on an important subclass of biophysical dynamics, the underlying approach generalizes to other physical or living systems that permit a mode representation while being  subject to exact or approximate  geometric constraints.
\par

 \begin{figure*}[t]
\includegraphics[width=17cm]{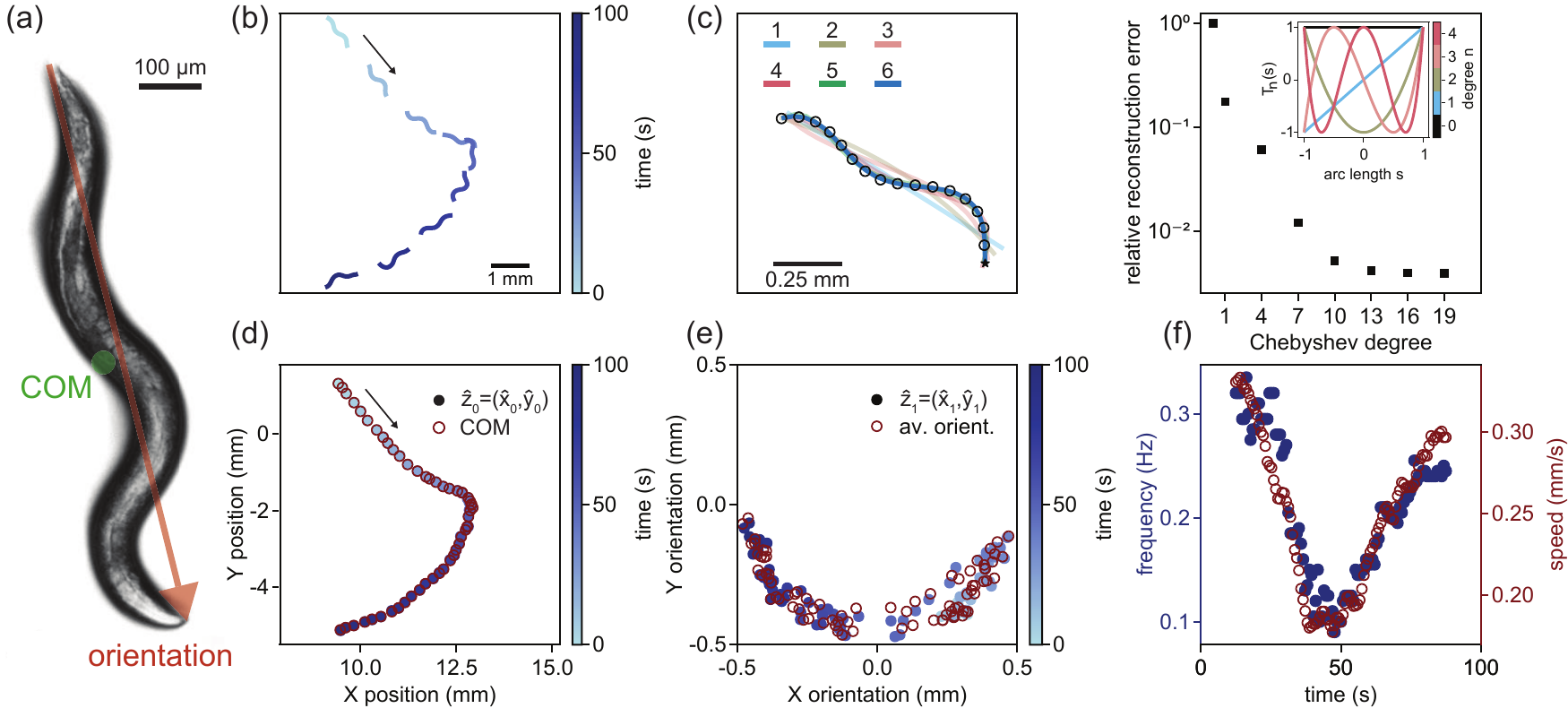}
\caption{\label{fig:fig_1} Chebyshev mode representation enables an efficient and interpretable low-dimensional description of undulatory locomotion across species and model systems. 
(a)~Experimental image of \emph{C. elegans} worm with center of mass (COM) and mean orientation overlayed. 
(b)~Tracked centerline of worm over 100 seconds. Arrow indicates direction of motion. 
(c)~A small number of Chebyshev polynomials suffices to accurately reconstruct the worm shape (left). Faint colored lines correspond to centerline reconstructions at different polynomial degrees. Reconstruction error (right) decays rapidly as the Chebyshev degree $n$ increases. 
(d)~The zeroth-order Chebyshev coefficients follow closely the worm's geometric COM,  illustrating the physical interpretability of the Chebyshev mode representation. 
(e)~Similarly, the first-order Chebyshev coefficients represent the tail-to-head worm orientation. 
(f)~The mode-averaged dominant frequency of Chebyshev mode oscillations correlates closely  with the locomotion speed of worm.}
\end{figure*}

The planar undulatory locomotion of an elongated worm-like object can be described by its centerline position in the complex plane $z(s, t) = x(s,t)+ i y(s,t)$, where $s \in[-1,1]$ is the arc length and $t$ denotes time~[Fig.~\ref{fig:fig_1}(a) and (b)]. While tens to hundreds of points are typically required for an accurate depiction of an organism's shape in position space~\cite{cermak2020whole}, interpretable lower-dimensional representations can often be obtained by projecting on suitable polynomial, trigonometric, or other basis functions~\cite{kantsler2012fluctuations,stephens2008dimensionality}. 
Although system-dependent representations, such as PCA-based eigenworms~\cite{stephens2008dimensionality, berman2014mapping}, yield near-optimal compression for a specific organism under fixed  experimental conditions, system-independent  orthogonal basis expansions enable  direct comparisons across different systems and experimental conditions~(SI). Moreover, system-dependent  representations are often non-differentiable making physically constrained modeling analytically intractable. Here, we use Chebyshev polynomials~\cite{boyd2001chebyshev} of the first kind, $T_k(s)$, which are known to have advantageous analytical and computational properties; in principle other basis functions could be chosen as well. The dynamics of the complex scalar field $z(s, t) = x(s,t) + iy(s,t)$ can then be represented in terms of its leading Chebyshev coefficients $\hat{z}_k(t)=\hat{x}_k(t)+i \hat{y}_k(t)$ up to degree $n$, defined by 
\begin{equation}
    \label{eq:cheb_coeffs}
    z(s, t) = \sum_{k = 0}^n T_k(s) \, \hat{z}_k(t).
\end{equation}
For the experimental imaging data analyzed below, ${n+1=10}$  modes suffice for achieving  reconstruction errors less than~$1\%$~[Fig.~\ref{fig:fig_1}(c); SI]. Since Chebyshev polynomials are orthogonal with respect to the weight function $w(s) =1/\sqrt{1-s^2}$~\cite{boyd2001chebyshev}, the  coefficients $\hat{z}_k$ are obtained by taking inner products,
\begin{equation}
 \hat{z}_k(t) = \frac{\gamma_k}{\pi}\int_{-1}^1 \mathrm{d}s\; w(s)\; T_k(s) \;z(s, t)
    \label{eq:transformation}
\end{equation}

\noindent where $\gamma_0 = 1$ and $\gamma_k = 2$ for $k > 0$. We illustrate the physical meaning of the Chebyshev modes using recent tracking microscopy video data~\cite{cermak2020whole} for  \emph{C.~elegans} [Fig.~\ref{fig:fig_1}(a) and (b)], a widely studied model organism with 95 body wall muscle cells, 302 neurons, and a rich set of behavioral states and corresponding locomotion patterns~\cite{flavell2020behavioral}. The real and imaginary parts of $\hat{z}_0(t)=\hat{x}_0(t)+i\hat{y}_0(t)$, obtained from Eq.~\eqref{eq:transformation} with~\mbox{$T_0(s)=1$}, describe the $w$-weighted Chebyshev center of mass (CCOM) of the moving worm, which follows closely the  geometric center of mass [Fig.~\ref{fig:fig_1}(d)]. The degree-$1$ coefficient $\hat{z}_1(t)$ with $T_1(s)=s$ represents the mean orientation of the worm [Fig.~\ref{fig:fig_1}(e), SI]. Similarly, the Chebyshev coefficients $\hat{z}_k$ of degree $k\ge 2$ encode curvature and higher deformation modes [Fig.~\ref{fig:fig_1}(c), inset]. The average dominant frequency across the mode oscillations closely matches the speed of the worm in real space [Fig.~\ref{fig:fig_1}(f);~SI].
\par
\begin{figure*}[t]
\includegraphics[width=17cm]{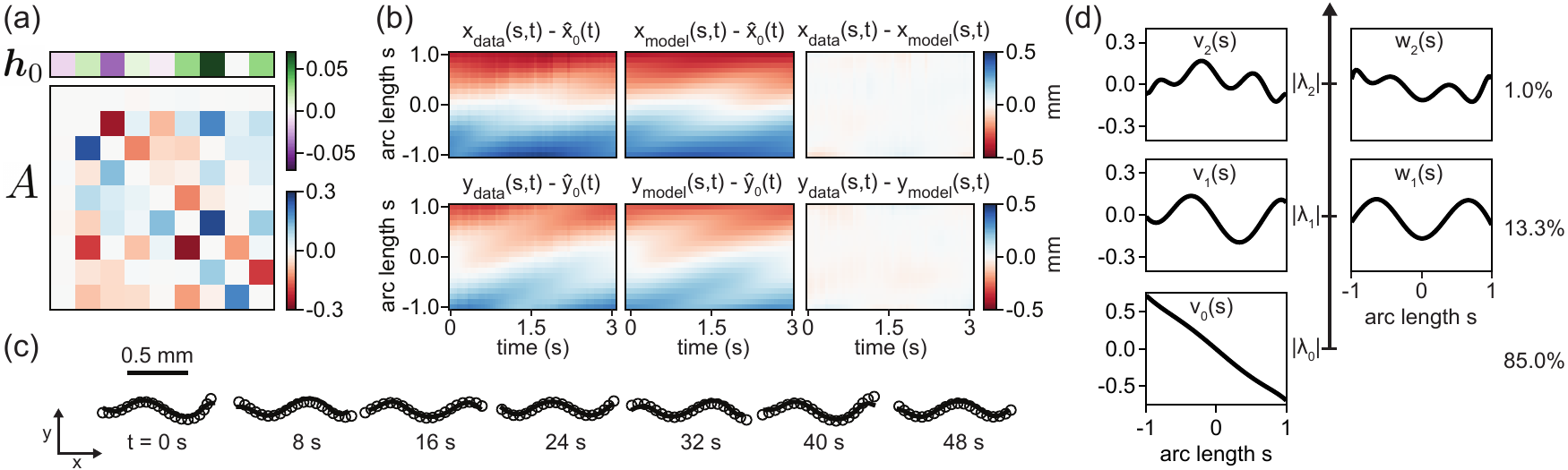}\textit{}
\caption{\label{fig:fig2} 
Inferred Schr\"odinger dynamics replicate stereotypical \emph{C. elegans} locomotion. 
(a)~Representative real propulsion vector $\boldsymbol{h}_0$ and Hamiltonian $H=S+iA$ for a minimal periodic straight-motion model [Eq.~\eqref{e:dynamics} with $S=0$ and equidistant spectrum of $H$] fitted to data from a single oscillation period ($\tau=3.05$~s). 
(b)~Kymographs of $x(s,t)$ and $y(s,t)$ coordinate fields for observed data (left) and model prediction (middle) show little deviation (right), confirming that Eq.~\eqref{e:dynamics} can accurately capture undulatory shape dynamics of \textit{C.~elegans}.
(c)~Real-space dynamics predicted by the Schr\"odinger model (line) is consistent with the observed worm dynamics (circles); see Movie~S1. Experimental data has been periodically extended for visualization to avoid overlapping body segments. 
(d)~Real-space shape functions [Eq.~\eqref{eq:rsFunctions}] corresponding to the three smallest magnitude eigenvalues, $\lambda^\pm_k = \pm k\lambda$ for $k=0,1,2$, account for $>98\%$ of the shape dynamics, enabling a generalizable low-rank description. More complex turning dynamics can be described using time-varying Hamiltonians with unconstrained spectra (Fig.~\ref{fig:fig4}; SI). 
}
\end{figure*}

Equipped with this representation, we seek to formulate a dynamical model for undulatory motion in mode space. Defining a combined mode vector ${\boldsymbol{\Psi}(t) =[\hat{z}_0, \ldots, \hat{z}_n]}\in \mathbb{C}^{n+1}$, the most general coupled linear first-order dynamics is  $\dot{\boldsymbol{\Psi}} = M \boldsymbol{\Psi}$. Note that the complex formulation is manifestly rotationally invariant, since a rotation by $\theta$ correspond to multiplication by~$e^{i\theta}$. Incorporating additional symmetries and invariances into the model imposes further structure on $M$. Translational invariance requires the CCOM $\boldsymbol{\psi}_0 = \hat{z}_0$ to decouple from the higher degree coefficients ${\boldsymbol{\hat{\psi}} = [\hat{z}_1, \ldots, \hat{z}_n]}\in \mathbb{C}^n$ that describe the orientation and shape (SI). Abbreviating $\partial_s z=\partial z/\partial s$, an additional biophysical constraint for undulatory motion is that the length of the centerline ${\ell(t) = \int_{-1}^1 \mathrm{d}s \,|\partial_s z|}$, remains approximately constant (SI). In mode space, length variations can be bounded  by conserving the convex quadratic functional 
\begin{equation}
\tilde{\ell}^2 = \int_{-1}^1 \mathrm{d}s \, |\partial_s z|^2 
= \boldsymbol{\hat\psi}^\dagger W \boldsymbol{\hat\psi}
\label{e:length_constraint}
\end{equation}
where $W$ is a symmetric matrix with elements $W_{k, m} = \int_{-1}^{1} \mathrm{d}s\, \partial_s T_k(s) \,\partial_s T_m(s)$. In particular, $W$~is positive definite and can thus be interpreted as a basis-specific metric. Taylor expanding the curve length $\ell$ around the space- and time-average of $|\partial_s z|^2$, denoted by $\langle \cdot \rangle$, shows that $\ell$ is approximately proportional to $\tilde{\ell}^2 / \sqrt{\langle |\partial_s z|^2 \rangle}$; additionally, the Cauchy-Schwarz inequality implies $\ell^2 \le 2\tilde{\ell}^2$ (SI). Therefore, demanding constant $\tilde{\ell}$ corresponds to an energetic penalty against contracting or lengthening, and ensures $\ell$ remains approximately constant and bounded. Keeping Eq.~\eqref{e:length_constraint} constant forces the shape-modes $\hat{\boldsymbol{\psi}}$ onto a hyperellipsoid, with axes determined by $W$. Using the Cholesky factorization $W = L L^\dagger$, this hyperellipsoid can be transformed to a unit hypersphere by defining the rescaled mode vector ${\boldsymbol{\psi} = (L^\dagger/\tilde{\ell})\boldsymbol{\hat{\psi}}}$. Under this transformation, the length constraint~\eqref{e:length_constraint} becomes a normalization condition
\begin{subequations}
\label{e:dynamics}
\begin{align}
\boldsymbol{\psi}^\dagger\boldsymbol{\psi} = 1.
\end{align}
Combined with rotational and translational invariance, the normalization restricts the class of permissible linear models to the form (SI)
\begin{align}
    \dot{\boldsymbol{\psi}_0} &= \boldsymbol{h}_0^\dagger\boldsymbol{\psi} \label{eq:COMdynamics}\\
    i\dot{\boldsymbol{\psi}} &= H\boldsymbol{\psi},
    \label{eq:SHAPEdynamics}
\end{align}
where $\boldsymbol{h}_0$ is a complex vector and $H$ is a complex Hermitian matrix with real eigenvalues. 
\end{subequations}
Equation~\eqref{eq:COMdynamics} describes how the CCOM dynamics couples to the body oscillations through $\boldsymbol{h}_0$. Equation~\eqref{eq:SHAPEdynamics}, which governs the shape dynamics, is mathematically equivalent to a Schr\"odinger equation with Hamiltonian $H$~\cite{schrodinger1926undulatory}. \par
\par
\begin{figure*}
\includegraphics[width=17cm]{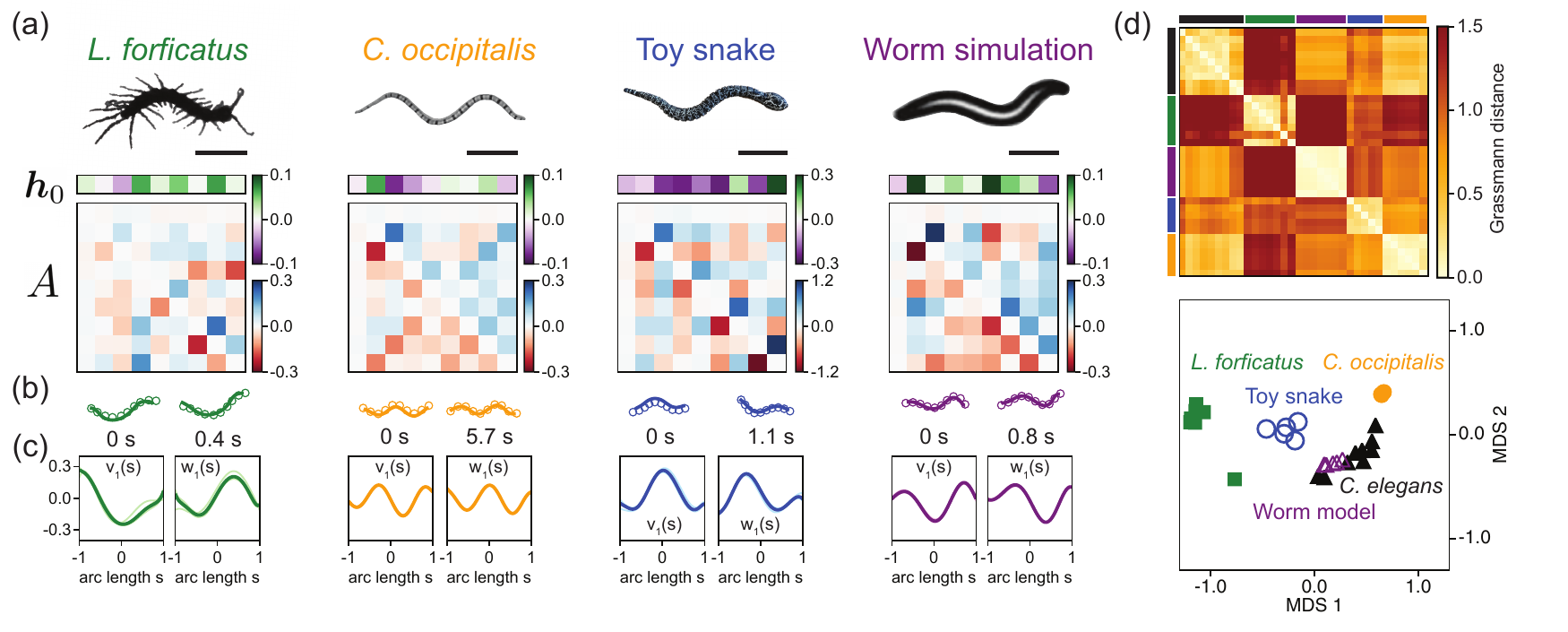}
\caption{\label{fig:fig3} Mode-space Hamiltonians provide a compact dynamical description of undulatory motion across different species and model systems. 
(a)~Living and nonliving systems~\cite{schiebel2019mechanical,boyle2012gait} analyzed here and representative  straight-motion Hamiltonians~$H=iA$ inferred from a single oscillation period. The eigenspaces of the Hamiltonians enable the comparison and classification of undulation dynamics in panel (d). Scale bars are 8 mm (centipede), 10 cm (snake), 10 cm (toy snake), 0.25 mm (worm model).
(b)~Inferred Schr\"odinger model dynamics (line) provide an accurate description of the observed dynamics (circles). Models were fitted on a single period $\tau=$ 0.19~s (centipede), 0.33~s (snake), 0.45~s (toy snake), 2.2~s (worm model);  see also Movie~S1. Experimental data has been periodically extended for visualization to avoid overlapping body segments.
(c)~The dominant shape eigenvectors $v_1(s)$ and $w_1(s)$  are consistent within each species and capture differences between  species.
(d)~Pairwise Grassmann distances between subspaces spanned by first excited eigenstates of the Hamiltonians (top) and its 2D planar embedding (bottom, constructed by a multidimensional scaling) capture the similarities and differences between undulatory locomotion in organisms, model simulations, and robots. Each point corresponds to a different trajectory. }
\end{figure*}
\begin{figure}[b]
\includegraphics{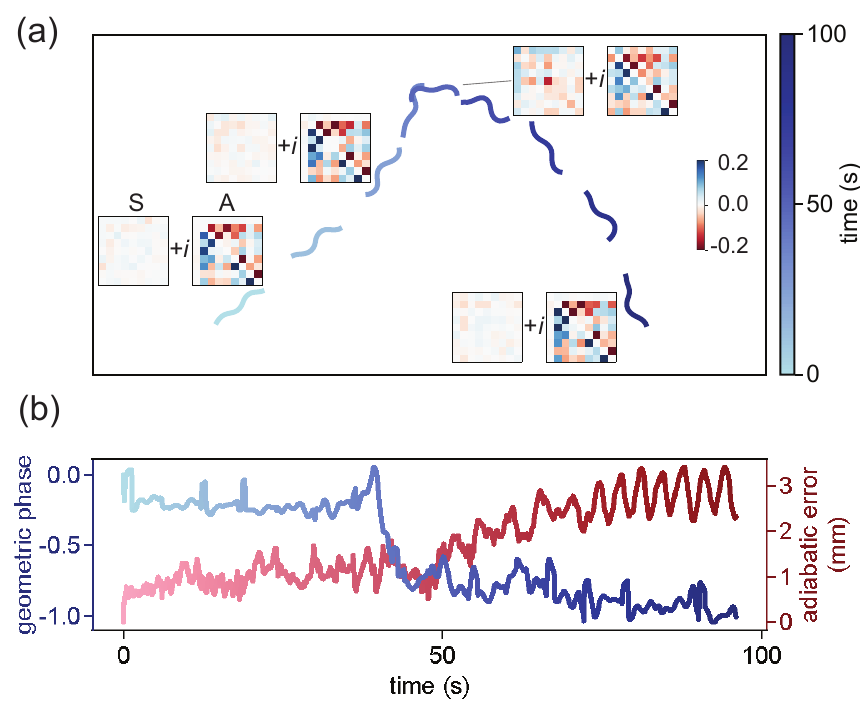}
\caption{\label{fig:fig4} 
Breakdown of adiabaticity during reversal turning behavior of \textit{C.~elegans}. 
(a) The turning part $S(t)$ of the Hamiltonian $H(t)=S(t)+i A(t)$ becomes switched on at the turn. (b) The turn is signaled by a sudden change in the geometric Berry phase (blue) of the dominant eigenvector (SI), and the RMS reconstruction error of the adiabatic approximation increases noticably after the turn (Movie~S2).}
\end{figure}

To confirm that Eqs.~\eqref{e:dynamics} can indeed describe and distinguish the undulatory dynamics of \textit{C.~elegans} worms~\cite{cermak2020whole} and other organisms and systems, we implemented an inference framework (SI) for estimating the propulsion vector $\boldsymbol{h}_0(t)$ and the shape Hamiltonian $H(t)$ from experimental data for short straight-motion segments (Figs.~\ref{fig:fig2},~\ref{fig:fig3}) as well as longer trajectories that include turning events~(Fig.~\ref{fig:fig4}). Before outlining the model inference procedure, recall that any Hermitian matrix $H$ can be decomposed in the form  $H = S + i A$, where $S$ is real symmetric and $A$ real skew-symmetric. In the present context, $S$ encodes turning behavior whereas $A$  governs straight locomotion: For straight motions,  $x$- and $y$-modes do not couple significantly, so that $\boldsymbol{h}_0$ is real and $S \approx  0$ and, hence, $H\approx iA$ in this case [Fig.~4(a); SI].
\par

Generally, both $\boldsymbol h_0$ and $H$ can be efficiently determined from tracked centerlines via a physics-informed dynamic mode decomposition~\cite{kutz2016dynamic, baddoo2021physics} that exploits matrix structure~\cite{sorber2015structured}.  
Since $H$ is Hermitian, it permits the decomposition $H=U\Lambda U^\dagger$, where $U$ is unitary and $\Lambda$ is a real diagonal matrix. This leaves $n^2$ parameters in $U$ and $\Lambda$ plus $2n$ in $\boldsymbol h_0$ to be estimated from data. If available data is limited, the number of parameters can be reduced further by imposing additional constraints on the spectrum of $H$ (SI). To avoid numerical differentiation of noisy data, our inference scheme compares numerically integrated predictions from  Eqs.~\eqref{e:dynamics} directly to the experimental data (SI). 
Our algorithm sequentially optimizes $U$, $\Lambda$ and $\boldsymbol{h}_0$ by minimizing deviations from both real space body shapes and mode space trajectories, to balance shape matching with model generalizability, and to prevent overfitting (SI).
Minimization is performed using gradient-based optimization~\cite{kingma2014adam,wright1999numerical,zhuang2020adabelief} with forward mode automatic differentiation through the ODE solver~\cite{rackauckas2020universal,romeo2021learning,ma2021comparison}. This scheme makes it possible to infer the instantaneous shape Hamiltonians $H(t)$ and the propulsion vectors $\boldsymbol{h}_0(t)$ from just a single oscillation period for straight motions (Figs.~\ref{fig:fig2} and~\ref{fig:fig3}) as well as from longer curved trajectories (Fig.~\ref{fig:fig4}). For \emph{C.~elegans} (Fig.~\ref{fig:fig2}) as well as for previously proposed neuro-mechanical worm models~\cite{boyle2012gait}, \textit{C.~occipitalis} snakes~\cite{schiebel2019mechanical}, snake robots, and \textit{L. forficatus} centipedes (Fig.~\ref{fig:fig3}), the best-fit straight-motion models based on Eqs.~\eqref{e:dynamics} with $H=iA$ accurately capture the undulatory dynamics (Movie~S1).

\par
Since the shape dynamics are encoded by the Hamiltonian $H$, we can use its eigenstates to compare and classify undulatory motion across species and  systems~\cite{griffiths2018introduction}. 
Indeed, for straight motions, it suffices to study the eigenstates of $A$. 
Considering $n=9$ as before, $A$ has one zero eigenvalue $\lambda_0 = 0$ corresponding to the zero-mode eigenvector $\boldsymbol{\phi}_0$, and $4$ distinct pairs of opposite sign eigenvalues $\lambda_{k\ge 1}^\pm$ with complex conjugate eigenvectors $\boldsymbol{\phi}_k^\pm$, where $\boldsymbol{\phi}_k^+ = (\boldsymbol{\phi}_k^-)^*$. 
We define two real orthogonal mode space vectors $\boldsymbol{v}_k = \Re(\boldsymbol{\phi}_k^+)$ and $\boldsymbol{w}_k = \Im(\boldsymbol{\phi}_k^+)$ that span the eigenspace of $\boldsymbol{\phi}_k^\pm$. 
The real space shape functions corresponding to the real mode space vectors are 
\begin{equation} \label{eq:rsFunctions}
   v_k(s) = \ell (L^{-1}\mathbf{T}(s))^\dagger \boldsymbol{v}_k, \,\quad w_k(s) = \ell (L^{-1}\mathbf{T}(s))^\dagger \boldsymbol{w}_k,
\end{equation}
where $\mathbf{T}(s) = [T_1(s), T_2(s), \dots, T_n(s)]$ is a vector of Chebyshev functions.
Time varying linear combinations of $v_k(s)$ and $w_k(s)$ give the instantaneous centerline reconstruction (SI). 
We find that the zero-function $v_0(s)$ is close to the best fit straight line through the motion, accounting for~85\% of the time-averaged centerline reconstruction while most of the oscillations are accounted for by the first excited-states $v_1(s)$ and $w_1(s)$ corresponding to the smallest magnitude non-zero eigenvalues (13.3\%). 
Since most ($>98\%$) of the dynamics is captured by the zero-state and first excited states, one can in fact further reduce the complexity of the Schr\"odinger model, by approximating $A$ through its projection $\hat{A}$ on the eigenspaces corresponding to the first two distinct eigenvalues.
This additional low-rank approximation also further reduces the risk of overfitting and hence improves model generalizability, similar to sparsity promotion in other dynamical inference methods~\cite{brunton2016discovering}.
\par
The compact low-rank characterization of the undulatory shape dynamics makes it possible to compare the locomotion behaviors of \emph{C. elegans}, previously proposed neuro-mechanical worm models~\cite{boyle2012gait}, \textit{C.~occipitalis} snakes~\cite{schiebel2019mechanical}, robotic toy snakes, and centipedes, by measuring the Grassmann distance~\cite{ye2016schubert} between the dominant eigenspaces of $\hat{A}$. 
As most of the variation of the oscillatory dynamics is contained in the first excited-states $\boldsymbol{v}_1$ and $\boldsymbol{w}_1$, we determined the pairwise Grassmann distances between the eigenspaces spanned by $\boldsymbol{v}_1$ and  $\boldsymbol{w}_1$ for the various systems~(SI). 
Both the distance matrix and a corresponding 2D phase diagram constructed by multidimensional scaling reveal that the neuro-mechanical worm model~\cite{boyle2012gait} succeeds in reproducing key dynamical aspects of \emph{C. elegans} locomotion, whereas the robotic toy snake used in our experiments is equally far from real snake or worm locomotion~(Fig.~\ref{fig:fig3}d).  
\par
Beyond inter-species comparisons, the above framework enables us to characterize behavioral transitions by borrowing concepts from quantum mechanics, such as Berry phases and adiabatic approximations~\cite{berry1984quantal}. To illustrate this, we focus on a longer \textit{C. elegans} trajectory during which the worm performs a turn [Fig.~\ref{fig:fig4}(a)] after briefly reversing its motion due to a change in neuro-mechanical activity~\cite{gray2005circuit}. By reconstructing the time-dependent Hamiltonian $H(t)=S(t)+i A(t)$ along the path (SI), we observe a significant increase in $||S(t)||$ at the turn whereas $A(t)$ remains approximately constant throughout. When the worm switches on $S$ to facilitate a turn, the instantaneous eigenvectors of $H(t)$ change (SI), signaled by a rapid change of the Berry phase [blue curve in Fig.~\ref{fig:fig4}(b)]. Furthermore, while the locomotion dynamics before the turn is well described by an adiabatic approximation (SI; Movie~S2), this approximation becomes inaccurate during the turn  [red curve in Fig.~\ref{fig:fig4}(b)].

\par
From a practical perspective, the above results show how symmetry-constrained mode representations can facilitate a low-dimensional description and efficient classification of biophysical dynamics.
The underlying inference framework is directly applicable to diagnose and quantify the effects of genetic or chemical perturbations on animal locomotion within and across species. 
From a theoretical perspective, the fact that translational and rotational invariance combined with a  quadratic integral constraint generically lead to a Schr\"odinger equation ~\cite{schrodinger1926undulatory} in mode space, promises advances in the quantitative understanding of biological systems, as the comprehensive toolbox of quantum physics~\cite{goldstein1992solitons, hasimoto1972soliton} now becomes available to characterize and predict behavioral dynamics.

\tocless \section{Acknowledgements}
A.E.C. and A.D.H. contributed equally to this work. This work was supported by the Department of Defense (DoD) through the National Defense Science and Engineering Graduate (NDSEG) Fellowship Program (A.E.C.), a MathWorks Science Fellowship (A.D.H.), the JPB Foundation (S.W.F.), a Sloan Research Fellowship (S.W.F.), a McKnight Foundation Scholar Award (S.W.F.), NSF Award 1845663 (S.W.F.), Sloan Foundation Grant G-2021-16758 (J.D.),  NSF Award DMS-1952706 (J.D.), and the Robert E. Collins Distinguished Scholarship Fund (J.D.).
We thank Daniel Goldman and Kelimar Diaz for providing the centipede movies.
\let\oldaddcontentsline\addcontentsline
\renewcommand{\addcontentsline}[3]{}
\bibliography{main}
\let\addcontentsline\oldaddcontentsline


\pagebreak
\widetext
\newpage
\begin{center}
\textbf{\large Supplemental Materials}
\end{center}
\setcounter{equation}{0}
\setcounter{figure}{0}
\setcounter{table}{0}
\setcounter{page}{1}
\makeatletter
\renewcommand{\theequation}{S\arabic{equation}}
\renewcommand{\thefigure}{S\arabic{figure}}
\renewcommand{\bibnumfmt}[1]{[S#1]}
\renewcommand{\citenumfont}[1]{S#1}

\tableofcontents

\section{Data preprocessing}\label{SI:data}
Raw data from the experiments consist of video files of the animal/simulation/robot motion. The analysis process for this data follows closely to that described in~\cite{cermak2020whole}. Each image frame is thresholded to obtain a binary image. These binary images are then dilated and eroded to fill in holes while preserving object shape. The resulting image is then thinned to obtain a centerline. Finally, a set number of equally spaced points along the centerline is generated.

To obtain the Chebyshev mode coefficients, we obtain $(x,y)$ positions at the Chebyshev points by linearly interpolating between the centerline points. We perform the transformation to a degree-19 Chebyshev polynomial, of which we use the mode coefficients up to degree-9 to represent the shape of the body.

In order to determine the dominant mode frequencies used in Figure~\ref{fig:fig_1}c, we take the Fourier transform of modes $\hat{x}_k$ and $\hat{y}_k$ for $k=1,...,n$. We extract the frequency which has the largest power in the spectra, which we call the dominant frequency. We then average the dominant frequency for all modes and plot this average in Figure~\ref{fig:fig_1}f.

\section{Toy snake experiments}
The toy snake was manufactured by Top Race and purchased from Amazon. The toy snake was placed on green construction paper on a hardwood floor and was remotely controlled to move in a straight line. Videos were filmed on an iPhone 12 Pro resting on a surface 1 meter above the floor. 
\begin{figure*}
    \centering
    \includegraphics[width=12cm]{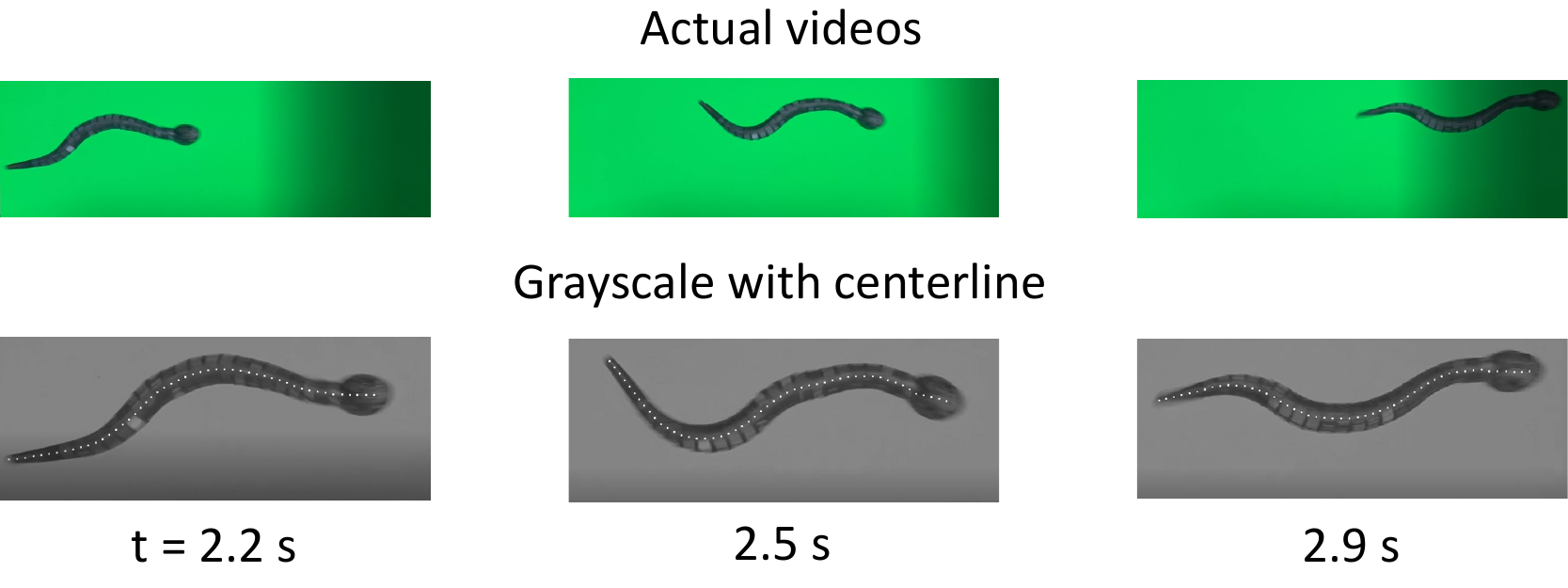}
    \caption{(top) Images from the toy snake video used in the paper. (bottom) Corresponding grayscale images with points along the centerline overlayed. 
    }
    \label{SIfig:toysnake_experiment}
\end{figure*}

\section{Reconstruction error}\label{SI:recon_error}
\begin{figure*}
    \centering
    \includegraphics{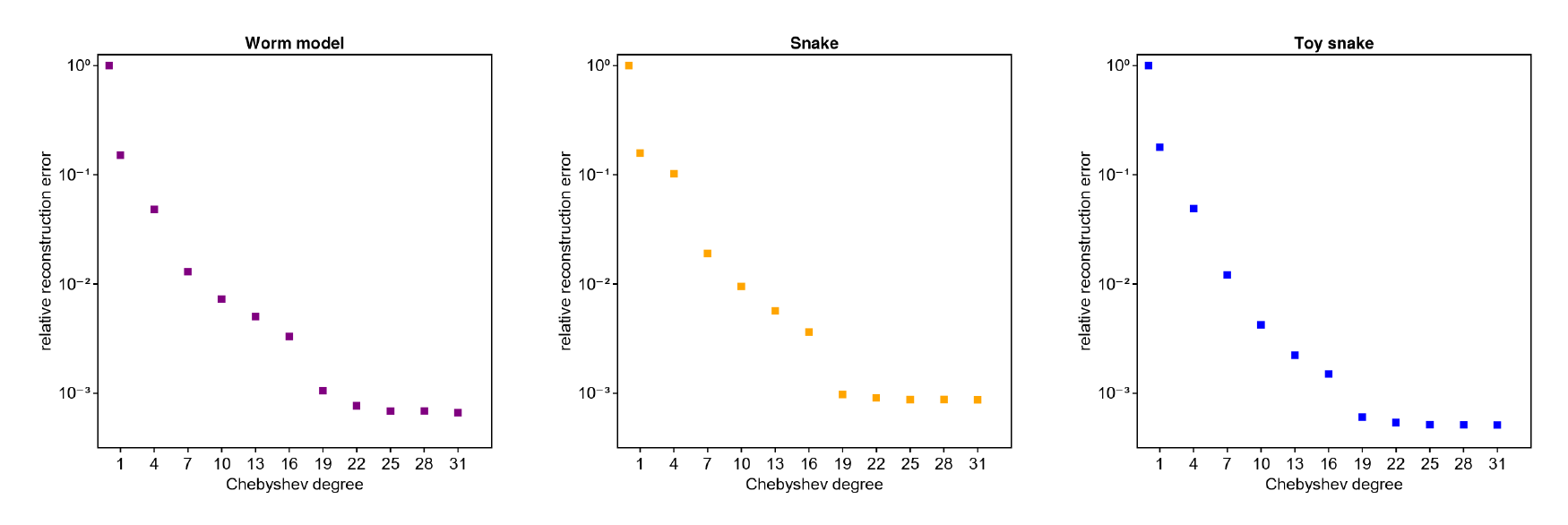}
    \caption{Reconstruction error plots for the worm simulation, snake, and toy snake. The same plot for the worm is shown in Main Text  Figure~\ref{fig:fig_1}(c).}
    \label{SIfig:recon_error}
\end{figure*}
The reconstruction error used in Figure~\ref{fig:fig_1}c and Figure~\ref{SIfig:recon_error} is calculated using
\begin{equation}
    \text{relative reconstruction error}=\frac{\sum_{t,s}\sqrt{(x_c-x_d)^2+(y_c-y_d)^2}}{\sum_{t,s}\sqrt{(\hat{x}_0-x_d)^2+(\hat{y}_0-y_d)^2}},
\end{equation}
where $x_c$ and $y_c$ are the $x$ and $y$ points calculated from the Chebyshev approximation and $x_d$ and $y_d$ are the $x$ and $y$ points from the experimental data. This can be interpreted as a relative mean absolute error, where we scale relative to the distances between all worm points from the worm center of mass. Therefore, the error is a measure of the distance deviations from the worm CCOM (a close approximation to the COM) accounted for by the Chebyshev polynomial approximation.

\subsection{System-specific versus general basis}

As mentioned in the main text, we use a prescribed general orthogonal basis system to represent animal shape to enable direct comparison across systems. 
This is in contrast to system-specific representations, such as PCA-based eigenworms.

\begin{figure*}[h]
    \centering
    \includegraphics[width=0.75\textwidth]{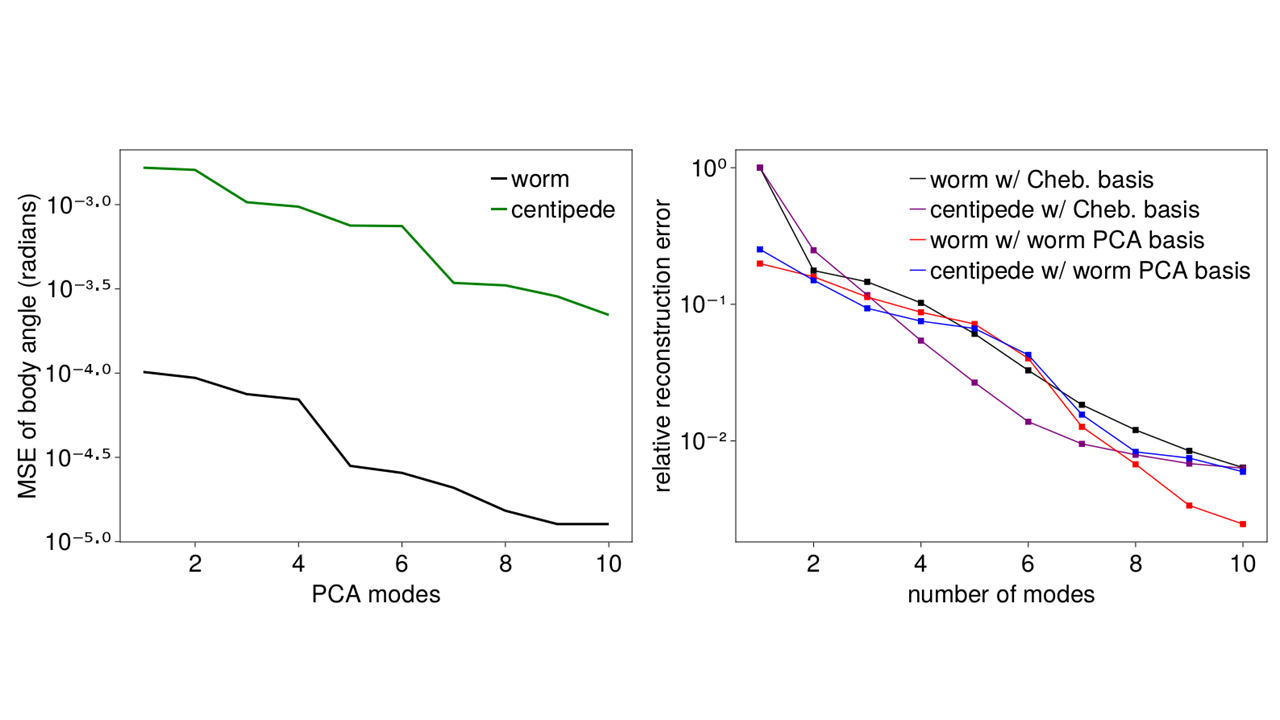}
    \caption{Reconstruction error of using different bases and representation. Left: Reconstruction error using the angle representation and eigenworm bases. While the basis is optimal for worms, the centipede has an order of magnitude larger error. Right: Reconstruction error using Chebyshev bases and a PCA basis in the $(x,y)$ representation.}
    \label{SIfig:eigenworm_reconstruction}
\end{figure*}

In Fig.~\ref{SIfig:eigenworm_reconstruction}, we compare the convergence results using the PCA-based basis in both the angle and $(x,y)$ representation. On the left, we show the approximation error using the same eigenworm basis (with the angle representation of posture) from~\cite{stephens2008dimensionality}. We compute the basis from the worm posture data and then apply this basis to the worm and centipede. We see that this basis reproduces worm posture well, but performs an order of magnitude worse for centipede posture. Using the eigenbasis calculated from the centipede would better reproduce the centipede posture, but then we would be unable to directly compare the resulting dynamical models in these bases. Therefore, we require a basis that can reconstruct well both species simultaneously to enable direct comparisons.
 
On the right, we compare the errors from the eigenbasis to the errors using the Chebyshev basis in the $(x,y)$ representation. We see that the reconstruction error from representing the worm with a Chebyshev basis converges at a similar rate to the reconstruction error from representing the worm with its PCA eigenbasis. When $10$ modes are used, as expected, the PCA basis achieves a lower reconstruction error as it is the optimal basis. If a reconstruction error of less than $1\%$ is desired then $8$ PCA basis modes would be required while $9$ Chebyshev modes would be required showing that both approaches produce comparable convergence rates. However, when we use the worm PCA basis on the centipede data, the reconstruction error initially converges slower than using the Chebyshev basis on the centipede, and for large modes the two reconstruction errors approach similar values. Again, if a reconstruction error of less than $1\%$ is required across both species $8$ PCA basis modes would be required while $9$ Chebyshev modes would be required. The need to keep an extra mode is outweighed, in our application, by the fact that Chebyshev polynomials provide continuous, differentiable modes that we can use in analytic calculations and provide coefficients that are interpretable as weighted moments of the curve shape. An additional benefit of working with prescribed basis functions is the ability to incorporate data that are sampled on different grids as the modes can be evaluated at any points within the domain.

\section{Interpretation of Chebyshev coefficients}
Starting from the definition of the Chebyshev modes Eq.~\eqref{eq:cheb_coeffs} we can express the $n = 1$ coefficients as, 
\begin{equation*}
    \begin{bmatrix}
    \hat{x}_1(t)\\
    \hat{y}_1(t)
    \end{bmatrix}
    = \frac{2}{\pi}\int_{-1}^1 \mathrm{d}s\; w(s)\; s\begin{bmatrix}
    x(s,t)\\
    y(s,t)
    \end{bmatrix} = \frac{2}{\pi}\int_{-1}^1 \mathrm{d}s\;\frac{1}{w(s)}\; \begin{bmatrix}
    x_s(s,t)\\
    y_s(s,t)
    \end{bmatrix}
\end{equation*}
where we use the fact that $T_1(s) = s$, $\int \mathrm{d}s\; s w(s) = -(1 - s^2)^{1/2} = -1/w(s)$ and integration by parts. The resulting expression is the $1/w$-weighted Chebyshev orientation.

\section{Constrained linear models in mode space: Real formulation}\label{SI:constrained_models}

Starting from a vector field representation of a centerline, $\mathbf{r}=(x(t, s), y(t, s))$, where $s$ is the arclength along the centerline and $\mathbf{r}$ is the Cartesian coordinate of the centerline at position $s$, we show how physical constraints impose structure on a linear model in mode space. 

\subsection{Rotational invariance}\label{SI:rotational_invariance}

We want our model to be invariant under a rotation of the coordinate system given by
$$
R(\theta) = \begin{pmatrix} \cos \theta & -\sin \theta \\ \sin \theta & \cos \theta \end{pmatrix}.
$$
Applying $R$ to \eqref{eq:cheb_coeffs} tells us how the coefficients transform, 
\begin{subequations}
\begin{align*}
\begin{bmatrix} x'(s, t) \\ y'(s, t) \end{bmatrix} &= \sum_{k = 0}^n T_k(s)\,\begin{bmatrix} \hat{x}'_k(t) \\ \hat{y}'_k(t) \end{bmatrix} \nonumber \\
R \begin{bmatrix} x(s, t) \\ y(s, t) \end{bmatrix} &= \sum_{k = 0}^n T_k(s)\,R\begin{bmatrix} \hat{x}_k(t) \\ \hat{y}_k(t) \end{bmatrix} \nonumber
\intertext{which implies that,}
\begin{bmatrix} \hat{x}'_k(t) \\ \hat{y}'_k(t) \end{bmatrix} &= R\begin{bmatrix} \hat{x}_k(t) \\ \hat{y}_k(t) \end{bmatrix} \\
&= \begin{bmatrix}  \hat{x}_k(t) \cos \theta - \hat{y}_k(t) \sin \theta \\ \hat{x}_k(t) \sin \theta + \hat{y}_k(t) \cos \theta \end{bmatrix}.
\end{align*}
\end{subequations}
Hence, the coefficient state vector $\boldsymbol{\Psi}$ transforms as $\boldsymbol{\Psi}' = (R \otimes \mathbb{I}_{n + 1}) \boldsymbol{\Psi}$ where $\otimes$ denotes the Kronecker product of two matrices and $\mathbb{I}_{n}$ is the $n \times n$ identity matrix. Applying this to a general linear dynamics for $\boldsymbol{\Psi}$, yields, 
\begin{subequations}
\begin{align}
    \dot{\boldsymbol{\Psi}'} &= M \boldsymbol{\Psi}' \label{SIeq:general_dynamics_prime} \\
    (R \otimes \mathbb{I}_{n + 1}) \dot{\boldsymbol{\Psi}} &= M (R \otimes \mathbb{I}_{n + 1}) \boldsymbol{\Psi} \label{SIeq:general_dynamics}
    \intertext{equating $M$ in ~\eqref{SIeq:general_dynamics_prime} to~\eqref{SIeq:general_dynamics} then gives the condition on $M$,}
    (R \otimes \mathbb{I}_{n + 1}) M &= M(R \otimes \mathbb{I}_{n + 1}) \label{SIeq:M_condition}
\end{align}
\end{subequations}
Writing~\eqref{SIeq:M_condition} out in block matrix form
\begin{equation}
    \begin{bmatrix} \mathbb{I}_{n + 1}\cos \theta & -\mathbb{I}_{n + 1}\sin \theta  \\ \mathbb{I}_{n + 1}\sin \theta & \mathbb{I}_{n + 1}\cos \theta\end{bmatrix} \begin{bmatrix} M_{xx} & M_{xy} \\ M_{yx} & M_{yy} \end{bmatrix} =
    \begin{bmatrix} M_{xx} & M_{xy} \\ M_{yx} & M_{yy} \end{bmatrix}
    \begin{bmatrix} \mathbb{I}_{n + 1}\cos \theta & -\mathbb{I}_{n + 1}\sin \theta  \\ \mathbb{I}_{n + 1}\sin \theta & \mathbb{I}_{n + 1}\cos \theta\end{bmatrix} \nonumber
\end{equation}
and comparing the left and right hand sides, yields the following two constraints,
$- M_{yx} = M_{xy} = \tilde{M}_2$ and $M_{xx} = \tilde{M}_{yy} = 
M_1$. Rotational invariance therefore enforces that $M$ has the following block structure, 
\begin{equation}
    \label{SIeq:M_rotations}
    M = \begin{bmatrix}
    \tilde{M}_1 & \tilde{M}_2 \\ -\tilde{M}_2 & \tilde{M}_1
    \end{bmatrix}  .
\end{equation}

\subsection{Translational invariance}\label{SI:translational_invariance}

We further expect that the model we learn should not depend on the origin of the coordinate system. The dynamics, therefore, should be invariant under $x' = x + c_x$ and $y' = y + c_y$. The coefficients in mode space transform as, $\hat{x}'_0 = \hat{x}_0 + c_x$, $\hat{y}'_0 = \hat{y}_0 + c_y$ and the other coefficients are unchanged. As a result, the coefficient vector $\boldsymbol{\Psi}$ transforms as, 
\begin{equation}
\boldsymbol{\Psi}' = \boldsymbol{\Psi} + \begin{bmatrix} c_x \boldsymbol{e}_1 \\ c_y \boldsymbol{e}_1 \end{bmatrix},
\end{equation}
where $\boldsymbol{e}_1$ is the standard $n + 1$ dimensional unit vector along the first dimension. The transformed dynamical equation using $M$ from~\eqref{SIeq:M_rotations} can be written as, 
\begin{align*}
\dot{\boldsymbol{\Psi}}' &= M \boldsymbol{\Psi}' \\
\frac{\mathrm{d}}{\mathrm{d}t} \begin{bmatrix} \boldsymbol{\Psi}_x + c_x \boldsymbol{e}_1 \\ \boldsymbol{\Psi}_y + c_y \boldsymbol{e}_1 \end{bmatrix} &= \begin{bmatrix} \tilde{M}_1 & \tilde{M}_2 \\ -\tilde{M}_2 & \tilde{M}_1 \end{bmatrix} \begin{bmatrix} \boldsymbol{\Psi}_x + c_x \boldsymbol{e}_1 \\ \boldsymbol{\Psi}_y + c_y \boldsymbol{e}_1\end{bmatrix} \\
\frac{\mathrm{d}}{\mathrm{d}t} \begin{bmatrix} \boldsymbol{\Psi}_x\\ \boldsymbol{\Psi}_y\end{bmatrix} &= \begin{bmatrix} \tilde{M}_1 & \tilde{M}_2 \\ -\tilde{M}_2 & \tilde{M}_1 \end{bmatrix} \begin{bmatrix} \boldsymbol{\Psi}_x\\ \boldsymbol{\Psi}_y\end{bmatrix} + 
\begin{bmatrix} \tilde{M}_1 & \tilde{M}_2 \\ -\tilde{M}_2 & \tilde{M}_1 \end{bmatrix} \begin{bmatrix}  c_x \boldsymbol{e}_1 \\ c_y \boldsymbol{e}_1\end{bmatrix}.
\end{align*}
For the dynamics to be invariant under arbitrary shifts $[c_x, c_y]$ we need $\tilde{M}_1 \boldsymbol{e}_1 = \boldsymbol{0}$ and $\tilde{M}_2 \boldsymbol{e}_1 = \boldsymbol{0}$, which implies that that the first columns of $\tilde{M}_1$ and $\tilde{M}_2$ must be all $0$. This decouples the dynamics of the $k = 0$ modes from the rest of the modes. Splitting $\boldsymbol{\Psi}$ into a $0$ mode vector $\boldsymbol{\psi}_0 = [\hat{x}_0, \hat{y}_0]$ and a higher mode vector $\boldsymbol{\hat{\psi}} = [\hat{x}_1, \ldots, \hat{x}_n, \hat{y}_1, \ldots, \hat{y}_n]$, the dynamics becomes, 
\begin{subequations}
\label{SIeq:M_translation}
\begin{align}
\dot{\boldsymbol{\psi}_0} &= 
\begin{bmatrix} \hat{\boldsymbol{m}}_1^\dagger & \hat{\boldsymbol{m}}_2^\dagger\\ -\hat{\boldsymbol{m}}_2^\dagger & \hat{\boldsymbol{m}}_1^\dagger \end{bmatrix}  \boldsymbol{\hat{\psi}} \\
\dot{\boldsymbol{\hat{\psi}}} &= \begin{bmatrix} \hat{M}_1 & \hat{M}_2 \\ -\hat{M}_2 & \hat{M}_1\end{bmatrix} \boldsymbol{\hat{\psi}}
\end{align}
\end{subequations}
with the same structure as before but new block elements.

\subsection{Length constraint}\label{SI:length_constraint}
For undulatory motion we have one additional constraint: the length of the centerline is approximately constant. The length of the centerline in terms of the real fields $x(s, t)$ and $y(s, t)$ is, \begin{equation}
    \label{SIeq:length_def}
    \ell(t) = \int_{-1}^1 \mathrm{d}s \, \sqrt{x_s(s, t)^2 + y_s(s, t)^2}.
\end{equation}
To allow for a convenient representation of an approximate length constraint in mode space we consider $\ell^2$. Using the Cauchy-Schwarz inequality, $\langle f, g \rangle^2 \le \langle f, f \rangle \langle g, g \rangle$, with metric $\langle f, g \rangle = \int_{-1}^1\mathrm{d}s\, f\cdot g$, we can derive a convex upper-bound on the square length using $f=1$ and $g=\sqrt{x_s(s, t)^2 + y_s(s, t)^2}$, 
\begin{align*}
\ell^2 &= \left(\int_{-1}^1 \mathrm{d}s \, 1 \cdot \sqrt{x_s(s, t)^2 + y_s(s, t)^2} \right)^2 \nonumber\\
&\le \left(\int_{-1}^1 \mathrm{d}s \, 1^2 \right) \left(\int_{-1}^1 \mathrm{d}s \,[ x_s(s, t)^2 + y_s(s, t)^2] \right) \nonumber\\
\ell^2 &\le 2 \tilde{\ell}^2
\end{align*}
where we define a convex approximate square length $\tilde{\ell}^2$, 
\begin{equation*}
    \tilde{\ell}^2 = \int_{-1}^1 \mathrm{d}s \, [x_s(s, t)^2 + y_s(s, t)^2].
\end{equation*}

\begin{figure*}
    \centering
    \includegraphics{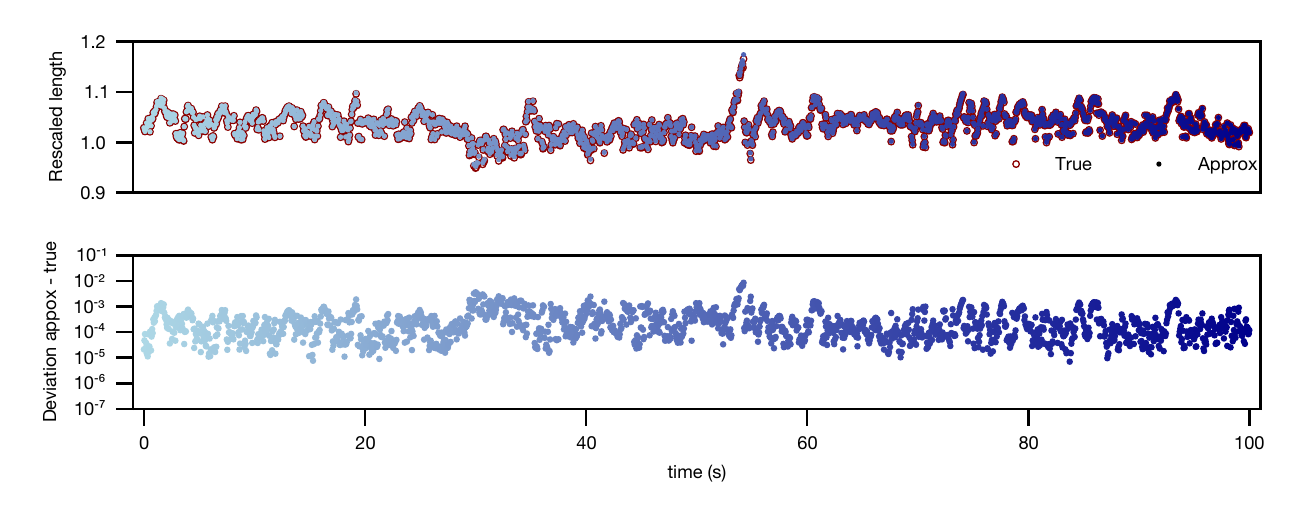}
    \caption{(top) True length $\ell$ and approximate length $\ell_a$ calculated using~\eqref{SIeq:l_approx} show close agreement. (bottom) Deviation between $\ell_a - \ell$ shows that the true deviation is much lower than the bound $0.13$. Also the deviation is always positive which means $\ell_a$ provides a close upper bound on $\ell$. 
    }
    \label{SIfig:length_approx}
\end{figure*}

We can find an approximation for $\ell$ in terms of $\tilde{\ell}$ by considering the Taylor expansion of $f(a, b) = \sqrt{a + b}$ around $a_0$ and $b_0$. Any $k = n + m$th order derivative of $f(a, b)$, is given by,
\begin{equation}
    \frac{\partial^n}{\partial a^n}\frac{\partial^m}{\partial b^m} f(a, b) = -\frac{(-1)^{n + m}}{2\sqrt{\pi}} \Gamma(n + m - 1/2) \frac{1}{(a + b)^{n + m - 1/2}} \nonumber
\end{equation}
The Taylor series then becomes,
\begin{align}
     \sqrt{a + b} &= \sqrt{a_0 + b_0} + \frac{1}{2 \sqrt{a_0 + b_0}} ((a + b) - (a_0 + b_0)) - \sum_{k = 2}^\infty \frac{(-1)^k \Gamma(k - 1/2)}{2\sqrt{\pi (a_0 + b_0)} k!} \left(\frac{a + b}{a_0 + b_0} - 1\right)^k \nonumber \\
     &= \frac{1}{2}\sqrt{a_0 + b_0} + \frac{a + b}{2 \sqrt{a_0 + b_0}}  - \sum_{k = 2}^\infty \frac{(-1)^k \Gamma(k - 1/2)}{2\sqrt{\pi (a_0 + b_0)} k!} \left(\frac{a + b}{a_0 + b_0} - 1\right)^k. \label{SIeq:general_sqrt_expan}
\end{align}
Setting $a = x_s^2$ and $b = y_s^2$ we can expand around $a_0 = \langle x_s^2 \rangle$ and $b_0 = \langle y_s^2 \rangle$ in~\eqref{SIeq:general_sqrt_expan}, where we use $\langle \cdot \rangle$ to represent the average value over $s$ and $t$
\begin{equation}
\langle f \rangle = \frac{1}{2T} \int_{0}^T \mathrm{d}t \int_{-1}^1 \mathrm{d}s \, f(s, t), \nonumber
\end{equation}
to get an expansion of the square-root term in~\eqref{SIeq:length_def}, 
\begin{equation}
\label{SIeq:sqrt_expansion}
    \sqrt{x_s^2 + y_s^2} = \frac{1}{2}\sqrt{\langle x_s^2 \rangle + \langle y_s^2 \rangle} + \frac{x_s^2 + y_s^2}{2 \sqrt{{\langle x_s^2 \rangle + \langle y_s^2 \rangle}}}  - \sum_{k = 2}^\infty \frac{(-1)^k \Gamma(k - 1/2)}{2\sqrt{\pi ({\langle x_s^2 \rangle + \langle y_s^2 \rangle})} k!} \left(\frac{x_s^2 + y_s^2}{{\langle x_s^2 \rangle + \langle y_s^2 \rangle}} - 1\right)^k
\end{equation}
Integrating~\eqref{SIeq:sqrt_expansion} over $s$, we get the following relationship between $\ell$ and $\tilde{\ell}$, 
\begin{align}
\ell = \int_{-1}^1 \mathrm{d}s \, \sqrt{x_s^2 + y_s^2} &= \sqrt{\langle x_s^2 \rangle + \langle y_s^2 \rangle} + \frac{1}{2 \sqrt{\langle x_s^2 \rangle + \langle y_s^2 \rangle}} \int_{-1}^1 \mathrm{d}s\, \left( x_s^2 + y_s^2 \right) + R(\Delta)\nonumber \\
\ell \approx \ell_a &= \sqrt{\langle x_s^2 \rangle + \langle y_s^2 \rangle} + \frac{1}{2 \sqrt{\langle x_s^2 \rangle + \langle y_s^2 \rangle}} \tilde{\ell}^2
\label{SIeq:l_approx}
\end{align}
where we define 
$$
\Delta = \frac{x_s^2 + y_s^2}{{\langle x_s^2 \rangle + \langle y_s^2 \rangle}} - 1
$$
a measure of how much the deviations vary from their average. We can get a bound for the magnitude of $R(\Delta)$ by evaluating the remaining summation and utilizing the triangle inequality, 
\begin{align}
    \lvert R(\Delta) \rvert &= \left\lvert \int_{-1}^{1} \mathrm{d}s\,\sum_{k = 2}^\infty \frac{(-1)^k \Gamma(k - 1/2)}{2\sqrt{\pi ({\langle x_s^2 \rangle + \langle y_s^2 \rangle})} k!} \Delta^k \right\rvert \nonumber \\
    &\le \int_{-1}^{1} \mathrm{d}s\,\sum_{k = 2}^\infty \frac{\Gamma(k - 1/2)}{2\sqrt{\pi ({\langle x_s^2 \rangle + \langle y_s^2 \rangle})} k!} \lvert \Delta \rvert^k \nonumber \\
    &\le \int_{-1}^{1} \mathrm{d}s\,\sum_{k = 2}^\infty \frac{\Gamma(k - 1/2)}{2\sqrt{\pi ({\langle x_s^2 \rangle + \langle y_s^2 \rangle})} k!} \Delta_M^k \nonumber \\
    &= \frac{1}{\sqrt{\pi ({\langle x_s^2 \rangle + \langle y_s^2 \rangle})}}\sum_{k = 2}^\infty \frac{\Gamma(k - 1/2)}{k!} \Delta_M^k \nonumber \\
    &= \frac{2 - \Delta_M - 2 \sqrt{1 - \Delta_M}}{\sqrt{\langle x_s^2 \rangle + \langle y_s^2 \rangle}}  \label{SIeq:l_approx_err} \nonumber
\end{align}
provided that $\Delta_M \le 1$, where $\Delta_M = \max_{s, t} \lvert\Delta\rvert$. For example, for the worm data considered here, $\Delta_M = 0.48$ and $\sqrt{\langle x_s^2 \rangle + \langle y_s^2 \rangle} = 0.50$, which give a value of $\lvert R(\Delta)\rvert< 0.15$, resulting in very close agreement between the true value of $\ell$ and the approximation calculated using~\eqref{SIeq:l_approx} (Fig.~\ref{SIfig:length_approx}). In practice the maximum error is much lower: the maximum deviation between the approximation and true value of $\ell$ for the worm data is $0.0084$ with a corresponding maximum relative error of $0.0073$. Unitless values of $\Delta_M$, $\sqrt{\langle x_s^2 \rangle + \langle y_s^2 \rangle}$, $\lvert R(\Delta)\rvert$ bounds and the maximum calculated errors are shown in Table~\ref{SItab:length_params}, further highlighting the validity of this approximation across of all the experimental systems studied here. 
\begin{table}[b]
    \centering
    \begin{tabular}{c|c|c|c|c}
         System & $\Delta_M$ & $\sqrt{\langle x_s^2 \rangle + \langle y_s^2 \rangle}/\ell$ & $\lvert R(\Delta)\rvert\times\ell$ bound & Maximum relative error \\
        \hline
        \hline
         \textit{C. elegans} &  0.48 & 0.50 & 0.15 & 0.0073 \\
         \hline
         Neuro-mechanical worm & 0.42 & 0.50 & 0.12 & 0.0064\\
         \hline
         \textit{C. occipitalis} & 0.40 & 0.50 & 0.10 & 0.0066 \\
         \hline
         Toy snake & 0.41 & 0.50 & 0.11 & 0.026 
    \end{tabular}
    \caption{Length approximation parameters for the systems studied here. The low relative errors highlight the validity of our relaxed length constraint. }
    \label{SItab:length_params}
\end{table}

Since $\ell_a$ is a function solely of $\tilde{\ell}^2$, keeping $\ell_a$ constant is the same as keeping $\tilde{\ell}^2$ constant. We, therefore, continue working under the assumption that $\tilde{\ell}^2$ is constant for undulatory motion.

\subsection{Length constraint in mode-space}

One of the benefits of working with $\tilde{\ell}^2$ rather than $\ell$ is that $x$ and $y$ appear quadratically, making it convenient to represent $\tilde{\ell}^2$ in mode space. We can express, $x_s(s, t)$ and $y_s(s, t)$ in mode space by differentiating~\eqref{eq:cheb_coeffs}, 
\begin{equation}
    \label{SIeq:s_diff_fields}
    \begin{bmatrix}
    x_s(s,t)\\
    y_s(s,t)
    \end{bmatrix} = \sum_{k=0}^n \, \frac{\mathrm{d}}{\mathrm{d}s} T_k(s) \begin{bmatrix}
    \hat{x}_k(t)\\
    \hat{y}_k(t)
    \end{bmatrix}.
\end{equation}
Substituting~\eqref{SIeq:s_diff_fields} into the expression for $\tilde{\ell}^2$~\eqref{e:length_constraint} gives
\begin{subequations}
\begin{align}
\tilde{\ell}^2 &= \int_{-1}^1 \mathrm{d}s\, \sum_{k, m = 1}^n \left[\hat{x}_k(t) \hat{x}_m(t) + \hat{y}_k(t) \hat{y}_m(t)\right] \frac{\mathrm{d}T_k}{\mathrm{d}s}(s) \frac{\mathrm{d}T_m}{\mathrm{d}s}(s) \nonumber\\
&= \sum_{k, m = 1}^n \left[\hat{x}_k(t) \hat{x}_m(t) + \hat{y}_k(t) \hat{y}_m(t)\right] \int_{-1}^1 \mathrm{d}s\,  \frac{\mathrm{d}T_k}{\mathrm{d}s}(s) \frac{\mathrm{d}T_m}{\mathrm{d}s}(s) \nonumber\\
&= \sum_{k, m = 1}^n \left[\hat{x}_k(t) \hat{x}_m(t) + \hat{y}_k(t) \hat{y}_m(t)\right] W_{k, m} \nonumber\\
&= \boldsymbol{\hat{\psi}}^\dagger \begin{bmatrix} W & 0 \\ 0 & W \end{bmatrix} \boldsymbol{\hat{\psi}}^\dagger
\end{align}
where we define the symmetric matrix $W$ with elements given by,  
\begin{equation}
    W_{k, m} = \int_{-1}^1 \mathrm{d}s\,  \frac{\mathrm{d}T_k}{\mathrm{d}s}(s) \frac{\mathrm{d}T_m}{\mathrm{d}s}(s). \label{SIeq:Wkn}
\end{equation}
\end{subequations}
Note that the values of $W$ are fixed by the choice of basis. Therefore, $W$ is a basis dependent constant. The matrix $W$ has several advantageous properties: it is symmetric positive definite, 
\begin{equation*}
\mathbf{a}^\dagger W \mathbf{a} = \sum_{k, m = 1}^n a_k W_{k, m} a_m = \int_{-1}^1 \mathrm{d}s\,  \left(\sum_{k = 1}^n a_k \frac{\mathrm{d}T_k}{\mathrm{d}s}(s)\right) \left(\sum_{m = 1}^n a_m \frac{\mathrm{d}T_m}{\mathrm{d}s}(s)\right) = \int_{-1}^1 \mathrm{d}s\,  \left(\sum_{k = 1}^n a_k \frac{\mathrm{d}T_k}{\mathrm{d}s}(s)\right)^2 \ge 0, 
\end{equation*}
since we can interpret the summation in the parentheses as the derivative of a Chebyshev series representation of some function $f(s) = \sum_{k = 0}^n a_k T_k(s)$. For the expression above to be $0$,  requires then that $f'(s) = 0$ for all $s$ and, therefore, $f(s) = a_0T_0(s)$, immediately implying that $\mathbf{a} = \boldsymbol{0}$ for equality in the above expression. 

\begin{figure}[h]
    \centering
    \includegraphics[width=0.25\textwidth]{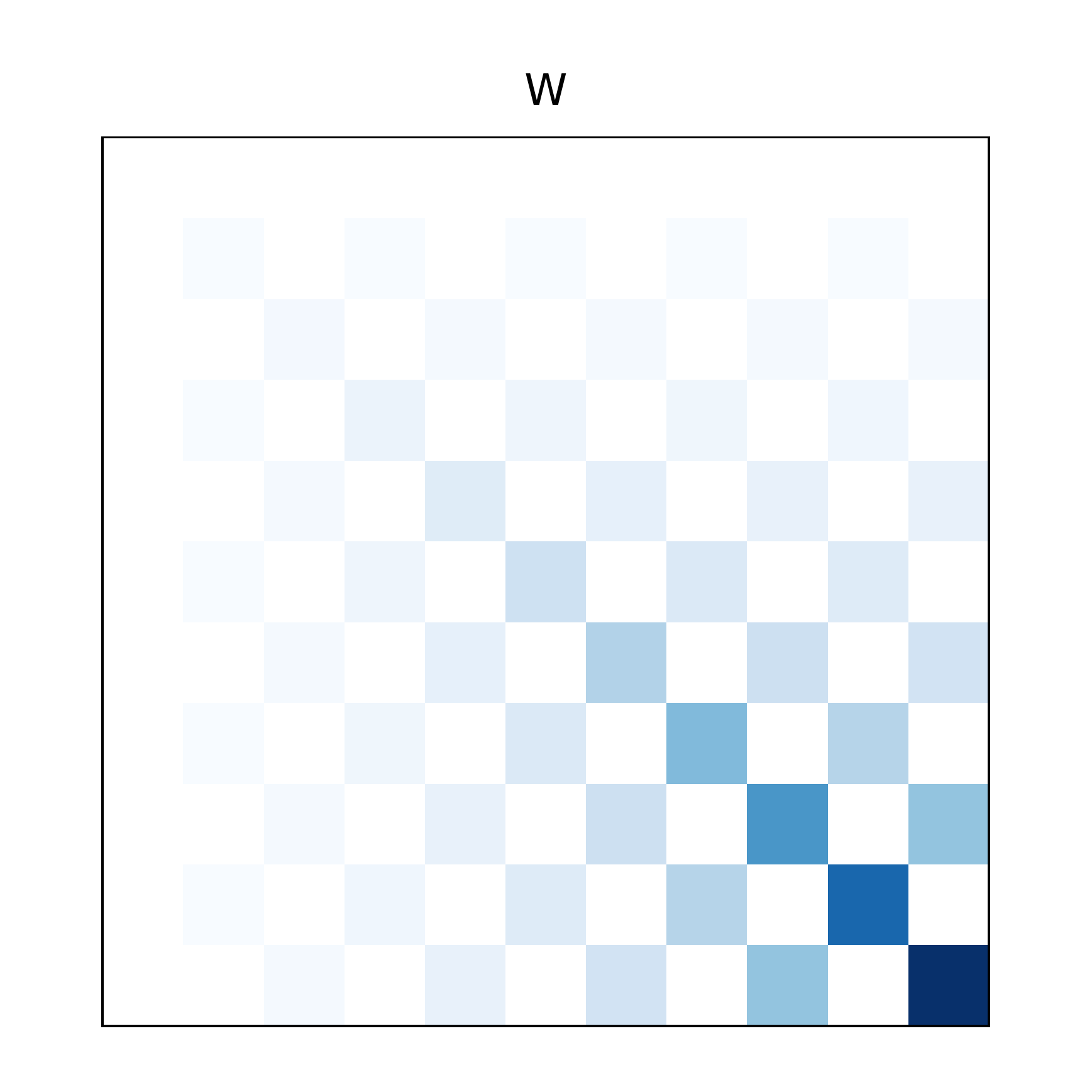}
    \caption{W matrix for $n = 10$}
    \label{SIfig:W}
\end{figure}

Making use of the following four Chebyshev polynomial identities, 
\begin{subequations}
\begin{align*}
    \frac{\mathrm{d}}{\mathrm{d} s} T_0(s) &= 0 \\ 
    \frac{\mathrm{d}}{\mathrm{d} s} T_k(s) &= k U_{k - 1}(s) \\   
    U_{n}(s) U_{m}(s) &= \sum_{k = 0}^{2\min(n, m)} U_{\lvert n -m \rvert + 2k}(s) \\
    \int_{-1}^1 \mathrm{d}{s}\, U_n(s) &= \left. \frac{T_{n +1}}{n+1}\right\rvert_{-1}^1 = \frac{1}{n + 1} - \frac{(-1)^{n+1}}{n + 1} = \frac{1 + (-1)^n}{n + 1}
\end{align*}
\end{subequations}
we can derive the values of $W_{m,n}$~\eqref{SIeq:Wkn}, 
\begin{align}
W_{m, n} &= \int_{-1}^1 \mathrm{d}{s}\, \frac{\mathrm{d}{T_n}}{\mathrm{d}s}(s) \frac{\mathrm{d}{T_m}}{\mathrm{d}s}(s) \nonumber\\
&= n m  \int_{-1}^1 \mathrm{d}{s}\, U_{n - 1}(s) U_{m - 1}(s) \nonumber\\
&= n m  \int_{-1}^1 \mathrm{d}{s}\,  \sum_{k = 0}^{\min(n - 1, m - 1)} U_{\left\lvert n - m \right\rvert + 2k}(s) \nonumber\\
&= n m   \sum_{k = 0}^{\min(n - 1, m - 1)}  \int_{-1}^1 \mathrm{d}{s}\, U_{\left\lvert n - m \right\rvert + 2k}(s) \nonumber\\
&= n m   \sum_{k = 0}^{\min(n - 1, m - 1)} \frac{1 + (-1)^{\left\lvert n - m \right\rvert + 2k}} {\left\lvert n - m \right\rvert + 2k +1} \nonumber\\ 
&= n m (1 + (-1)^{\left\lvert n - m \right\rvert} ) \sum_{k = 0}^{\min(n, m) - 1} \frac{1} {\left\lvert n - m \right\rvert + 2k +1} \nonumber \\
&= \begin{cases} 0 & \text{ if } m - n \text{ even} \\
2nm \sum_{k = \left\lvert n - m  \right\rvert + 1}^{n + m - 1} \frac{1} {k} &\text{ if } m - n \text{ even}\end{cases} \nonumber
\end{align}
which shows that $W$ has a checkerboard pattern and is diagonally dominant (Fig.~\ref{SIfig:W}).

The basis dependent matrix $W$ defines a hyperellipsoid that the mode vector $\boldsymbol{\hat{\psi}}$ lies on. Since $W$ is symmetric, positive-definite it has a Cholesky factorization $W = L L^\dagger$, which we can use to define a new mode vector
\begin{equation*}
    \boldsymbol{\psi} = \frac{1}{\tilde{\ell}} \begin{bmatrix} L^\dagger & 0 \\ 0 & L^\dagger  \end{bmatrix} \boldsymbol{\hat{\psi}}
\end{equation*}
that lies on the unit hypersphere. Since the matrix we apply is block-diagonal, applying this transformation to~\eqref{SIeq:M_translation} does not change the structure and we get the transformed equations, 
\begin{subequations}
\begin{align*}
    \dot{\boldsymbol{\psi}_0} &= \begin{bmatrix} \boldsymbol{{m}}_1^\dagger & \boldsymbol{{m}}_2^\dagger \\ -\boldsymbol{{m}}_2^\dagger & \boldsymbol{{m}}_1^\dagger \end{bmatrix} \boldsymbol{\psi} \\ 
    \dot{\boldsymbol{\psi}} &= \begin{bmatrix} M_1 & M_2 \\ - M_2 & M_1  \end{bmatrix} \boldsymbol{\psi}.
\end{align*}
Under this transformation the length constraint becomes,
\begin{equation*} 
    \boldsymbol{\psi}^\dagger \boldsymbol{\psi} = 1 .
\end{equation*}
\end{subequations}
Differentiating the new constraint with respect to time we get a dynamical constraint for length conservation, 
\begin{align*}
    \dot{\boldsymbol{\psi}}^\dagger \boldsymbol{\psi} + \boldsymbol{\psi}^\dagger \dot{\boldsymbol{\psi}} &= 0 \\
    \boldsymbol{\psi}^\dagger \begin{bmatrix} M_1^\dagger + M_1 & M_2 - M_2^\dagger \\ M_2^\dagger - M_2 & M_1 + M_1^\dagger \end{bmatrix} \boldsymbol{\psi} &= 0.
\end{align*}
For this to be true for any $\boldsymbol{\psi}$, we must further constrain the real block-diagonal $M_1 = - M_1^\dagger = A$ to be skew-symmetric and the real block-off-diagonal $M_2 = M_2^\dagger = S$ to be symmetric. The final real form of our constrained dynamics is then given by, 
\begin{subequations}
\label{SIeq:final_real_dynamics}
\begin{align}
    \dot{\boldsymbol{\psi}_0} &= \begin{bmatrix} \boldsymbol{{m}}_1^\dagger & \boldsymbol{{m}}_2^\dagger \\ -\boldsymbol{{m}}_2^\dagger & \boldsymbol{{m}}_1^\dagger \end{bmatrix} \boldsymbol{\psi} \\
    \dot{\boldsymbol{\psi}} &= \begin{bmatrix} A & S \\ - S & A \end{bmatrix} \boldsymbol{\psi} \label{SIeq:final_real_dynamics_shape}
\end{align}
\end{subequations}
The full dynamical matrix has now been constrained to skew-symmetric and must, therefore, have purley imaginery eigenvalues. Therefore, multiplying~\ref{SIeq:final_real_dynamics_shape} by the imaginary unit $i$, results in our final complex Schr{\"o}dinger type equations, that take the form of Schro{\"o}dinger's equations in quantum mechanics for the shape dynamics,
\begin{subequations}
\label{SIeq:final_complex_dynamics}
\begin{align}
    \dot{\boldsymbol{\psi}_0} &= \begin{bmatrix} \boldsymbol{{h}}_1^\dagger & \boldsymbol{{h}}_2^\dagger \\ -\boldsymbol{{h}}_2^\dagger & \boldsymbol{{h}}_1^\dagger \end{bmatrix} \boldsymbol{\psi} \\
    i\dot{\boldsymbol{\psi}} &= \begin{bmatrix} H_1 & H_2 \\ - H_2 & H_1 \end{bmatrix} \boldsymbol{\psi} = H\boldsymbol{\psi} \label{SIeq:final_complex_dynamics_shape}
\end{align}
\end{subequations}
when $H_1 = iA$ is a Hermitian matrix and $H_2 = iS$ is a skew-Hermitian matrix. 

\subsection{Straight motion}

For straight motion we expect that the modes associated with $x$ and the modes associated with $y$ do not interact significantly leading to the further simplification that $H_2 = S = 0$ in the final constrained dynamics.

\section{Constrained linear models in mode space: Complex formulation}\label{SI:constrained_models}

Starting from a vector field representation of the centerline, $\mathbf{r} = (x(t, s), y(t, s))$, where $s \in [-1, 1]$ is the parameter along the centerline and $\mathbf{r}$ is the Cartesian coordinate of the centerline at $s$, we can construct a single complex field $z(t, s) = x(t, s) + i y(t, s)$. We can represent this field in Chebyshev polynomials,
\begin{equation}
    z(t, s) = \sum_{k = 0}^n T_k(s) \hat{z}_k(t) = \sum_{k = 0}^n T_k(s) \left[\hat{x}_k(t) + i \hat{y}_k(t)\right]. \label{SIeq:zexpansion}
\end{equation}
We define the complex coefficient state vector $\boldsymbol{\Psi}_z = [\hat{z}_0, \, \hat{z}_1, \, \ldots, \hat{z}_n]$ from \eqref{SIeq:zexpansion} and consider a general linear dynamics $\dot{\boldsymbol{\Psi}}_z = M \boldsymbol{\Psi}_z$ for some complex matrix $M$.

Rotation of the coordinate system corresponds to multiplying $z$ by $e^{i\theta}$, $z' = e^{i\theta}z$, which implies that the coefficients transform as $\hat{z}'_k(t) = e^{i\theta}\hat{z}_k(t)$. Hence the complex coefficient state vector $\boldsymbol{\Psi}_z$ transforms as $\boldsymbol{\Psi}'_z = e^{i\theta}\boldsymbol{\Psi}_z$. Considering a general linear dynamics 
\begin{subequations}
\begin{align*}
    \dot{\boldsymbol{\Psi}}_z' &= M \boldsymbol{\Psi}_z' \\ 
    \implies e^{i\theta}\dot{\boldsymbol{\Psi}}_z &= M e^{i\theta} \boldsymbol{\Psi}_z \\
    \implies\dot{\boldsymbol{\Psi}}_z &= M \boldsymbol{\Psi}_z
\end{align*}
\end{subequations}
we see that using this construction the dynamics are automatically rotationally invariant.

Invariance under translation requires that the dynamics are invariant under the shift $z' = z + c_z$. Under this shift the coefficients transform as $\boldsymbol{\Psi}_z' = \boldsymbol{\Psi}_z + c_z \boldsymbol{e}_1$ and the transformed dynamics \begin{subequations}
\begin{align*}
    \dot{\boldsymbol{\Psi}}_z' &= M\left(\boldsymbol{\Psi}_z + c_z \boldsymbol{e}_1 \right)\\
    \implies \dot{\boldsymbol{\Psi}}_z &= M\boldsymbol{\Psi}_z + c_z M\boldsymbol{e}_1
\end{align*}
\end{subequations}
show that we need $M\boldsymbol{e}_1 = \boldsymbol{0}$. The first column of $M$ is zero decoupling the $\hat{z}_0$ dynamics from the higher mode dynamics of $\hat{\boldsymbol{\psi}}_z = [\hat{z}_1, \, \hat{z}_2, \, \ldots, \, \hat{z}_n]$, 
\begin{subequations} \label{SIeq:complex_decoupled}
    \begin{align}
        \dot{\hat{z}}_0 &= \boldsymbol{m}^\dagger \hat{\boldsymbol{\psi}}_z \\
        \dot{\hat{\boldsymbol{\psi}}}_z &= \hat{M} \hat{\boldsymbol{\psi}}_z.
    \end{align}
\end{subequations}

In the complex formulation the relaxed length constraint becomes,
\begin{subequations} 
\begin{align*}
    \tilde{\ell}^2 &= \int_{-1}^1 \mathrm{d}s \, (x_s^2 + y_s^2) = \int_{-1}^1 \mathrm{d}s \, \bar{z}_s z_s \\
    &= \sum_{n, m = 1}^n \bar{\hat{z}}_n \left[\int_{-1}^1 \mathrm{d}s \, \frac{\mathrm{d}T_n}{\mathrm{d}s}\frac{\mathrm{d}T_m}{\mathrm{d}s} \right] \hat{z}_m \\
    &= \hat{\boldsymbol{\psi}}_z^\dagger W \hat{\boldsymbol{\psi}}_z \\
\end{align*}
\end{subequations} 
where $W$ is the same as in \eqref{SIeq:Wkn}.
We use the Cholesky decomposition of $W = LL^\dagger$ to define the rescaled coefficients $\boldsymbol{\psi}_z = L^\dagger \hat{\boldsymbol{\psi}}_z / \ell$ under which the relaxed length constraint becomes
\begin{equation*}
    1 = \boldsymbol{\psi}_z^\dagger \boldsymbol{\psi}_z.
\end{equation*}
Under the rescaling the dynamics~\eqref{SIeq:complex_decoupled} becomes 
\begin{align*}
\dot{\boldsymbol{\psi}}_z = L^\dagger \dot{\hat{\boldsymbol{\psi}}}_z / \ell &= L^\dagger \hat{M} \hat{\boldsymbol{\psi}}_z/\ell \\
&= L^\dagger \hat{M} (L^\dagger)^{-1} \boldsymbol{\psi}_z \\
&= M \boldsymbol{\psi}_z.
\end{align*}
In order to satisfy the unit norm constraint we require that, 
\begin{equation}
\dot{1} = 0 = \dot{\boldsymbol{\psi}}_z^\dagger \boldsymbol{\psi}_z + \boldsymbol{\psi}_z^\dagger \dot{\boldsymbol{\psi}}_z = \boldsymbol{\psi}_z^\dagger (M^\dagger + M) \boldsymbol{\psi}_z
\end{equation}
giving the condition that $M = -M^\dagger$ is skew-Hermitian. Since $M$ is skew-Hermitian we can write it as $iH$ where $H$ is Hermitian yielding the following set of equations for the constrained complex dynamics of the center line, 
\begin{subequations}
    \begin{align}
        \boldsymbol{\psi}^\dagger_z \boldsymbol{\psi}_z &= 1 \\
        \boldsymbol{h}^\dagger \boldsymbol{\psi}_z &= \dot{\hat{z}}_0 \\
        i H \boldsymbol{\psi}_z &= \dot{\boldsymbol{\psi}}_z
    \end{align}
\end{subequations}

We note that we can write $M = A - iS$ where $A$ and $S$ are real skew-symmetric and symmetric matrices respectively. Then the dynamics become $M\boldsymbol{\psi}_z = (A - iS)(\boldsymbol{\psi}_x + i \boldsymbol{\psi}_y) = A\boldsymbol{\psi}_x + S\boldsymbol{\psi}_y + i(A\boldsymbol{\psi}_y - S \boldsymbol{\psi}_x)$. Taking real and imaginary components we get, $\dot{\boldsymbol{\psi}}_x = A\boldsymbol{\psi}_x + S\boldsymbol{\psi}_y$ and $\dot{\boldsymbol{\psi}}_y = -S\boldsymbol{\psi}_x + A\boldsymbol{\psi}_x$. These two equations can be summarized in a matrix equation, 
\begin{equation*}
    \begin{bmatrix}
    \dot{\boldsymbol{\psi}}_x \\ \dot{\boldsymbol{\psi}}_y
    \end{bmatrix} = 
    \begin{bmatrix}
    A & S \\ -S & A
    \end{bmatrix}
        \begin{bmatrix}
    \boldsymbol{\psi}_x \\ \boldsymbol{\psi}_y
    \end{bmatrix}
\end{equation*}
the same as \eqref{SIeq:final_real_dynamics_shape}.

\section{Model inference}\label{SI:model_inference}

\subsection{Model inference: theory}

The problem of learning an equation of the form~\eqref{SIeq:final_real_dynamics} from data for $\boldsymbol{\psi}$ and $\boldsymbol{\psi}_0$ at discrete time points $\{t_n\}_{n = 0}^M$ can be formulated as a physics-informed dynamic mode decomposition (PI-DMD) optimization problem. In continuous time the problem becomes,
\begin{equation}
    \min_A \sum_{n = 0}^T \lVert \dot{\boldsymbol{\psi}}(t_n) - A \boldsymbol{\psi}(t_n) \rVert_2^2 = \min_A \lVert \dot{P} - A P \rVert_F^2 \label{SIeq:cont_dmd}
\end{equation}
where $P = \left[\boldsymbol{\psi}(t_0) \,  \boldsymbol{\psi}(t_1) \, \boldsymbol{\psi}(t_2) \, \cdots \, \boldsymbol{\psi}(t_T)\right]$ is the matrix whose columns consists of the discrete time samples of $\boldsymbol{\psi}$. The minimization problem in~\eqref{SIeq:cont_dmd} has an analytical minimum in terms of the singular value decomposition (SVD) of $P$ but requires numerically differentiating noisy data to calculate $\dot{P}$ an ill-posed and challenging problem. We, therefore, formulate the problem in discrete time. The general solution of a linear ODE of the form, $\dot{\mathbf{x}} = M \mathbf{x}$, is $\mathbf{x}(t) = \exp(A t) \mathbf{x}(0)$. If the data are seperated by a constant time step $\Delta t$ we can reformulate~\eqref{SIeq:cont_dmd} in the form, 
\begin{equation}
    \min_A \sum_{n = 0}^{T-1} \left\lVert \boldsymbol{\psi}(t_{n + 1}) - e^{A\Delta t} \boldsymbol{\psi}(t_n) \right\rVert_2^2 = \min_A \left\lVert P_{2:T} - e^{A\Delta t} P_{1:T-1} \right\rVert_F^2 \label{SIeq:discrete_dmd}
\end{equation}
where $P_{1:T-1}$ consists of the first $T-1$ columns of $P$ and $P_{2:T}$ consists of the last $T-1$ columns. The skew-symmetric structure of the continuous time problem does not transfer to the discrete time problem, instead $\exp(A \Delta t)$ is an orthogonal matrix with a fixed form of its eigenvalues.
Additionally, there is no guarantee that the matrix $A$ will produce reintegrated trajectories close to the original input data. Exploiting the matrix exponential solution of linear ODEs we can modify~\eqref{SIeq:discrete_dmd} 
\begin{equation}
    \min_A \sum_{m = 1}^{T}  \sum_{n = 0}^N w_n\left[\boldsymbol{\psi}(t_{m}) - e^{A t_m} \boldsymbol{\psi}(t_0) \right]_n^2 \label{SIeq:optloss}
\end{equation}
where we introduce the possibility of a weighting function $w_n$ on the $n$th mode to account for the magnitude variations across the modes. The spectral theorem for skew-symmetric matrices tells us that a real $N \times N$ skew-symmetric matrix can be written in the form, 
\begin{equation*} \label{SIeq:spectral_decomp}
A = Q \Sigma Q^\dagger = \begin{bmatrix}
\vdots & \vdots & \vdots & \vdots &&\vdots & \vdots & \vdots \\
\boldsymbol{v}_1 & \boldsymbol{w}_1 & \boldsymbol{v}_2 & \boldsymbol{w}_2 & \cdots & \boldsymbol{v}_r & \boldsymbol{w}_r& \boldsymbol{v}_0 \\
\vdots & \vdots & \vdots & \vdots &&\vdots & \vdots & \vdots
\end{bmatrix} \begin{bmatrix} \Lambda_1  \\ & \Lambda_2 \\ && \ddots \\ &&& \Lambda_r \\ &&&& 0 \end{bmatrix} \begin{bmatrix}
\cdots & \boldsymbol{v}_1^\dagger & \cdots \\ \cdots & \boldsymbol{w}_1^\dagger & \cdots\\ \cdots & \boldsymbol{v}_2^\dagger & \cdots \\ \cdots & \boldsymbol{w}_2^\dagger & \cdots\\ &\vdots& \\ \cdots & \boldsymbol{v}_r^\dagger & \cdots \\ \cdots & \boldsymbol{w}_r^\dagger & \cdots\\ \cdots & \boldsymbol{v}_0^\dagger & \cdots
\end{bmatrix}
\end{equation*}
where $Q$ is a real orthogonal matrix $Q^\dagger Q = \mathbb{I}$, the $\Lambda_i$ are $2 \times 2$ blocks matrices $$\Lambda_i = \begin{bmatrix} 0 & \lambda_i \\ -\lambda_i & 0 \end{bmatrix} $$ with $r \le \lfloor N/2 \rfloor$ the number of distinct complex conjugate pairs of eigenvalues and the remaining block is a $N - 2r \times N - 2r$ zero matrix. When $N$ is odd there must always be a $0$ row and column in $\Sigma$. The matrix exponential then has the simple form, $\exp(At) = Q \exp(\Sigma t) Q^\dagger$, where
$$
\exp(\Sigma t) = \begin{bmatrix} \exp(\Lambda_1 t)  \\ & \exp(\Lambda_2 t) \\ && \ddots \\ &&& \exp(\Lambda_r t)\\ &&&& \mathbb{I} \end{bmatrix}
$$
and
$$
\exp(\Lambda_i t) = \begin{bmatrix} \cos \lambda_i t & \sin \lambda_i t \\ -\sin \lambda_i t & \cos \lambda_i t\end{bmatrix}  .
$$
The optimization loss function~\eqref{SIeq:optloss} can then be written as, 
\begin{equation}
    \min_{Q, \{\lambda_i\}_{i = 1}^r} \sum_{m = 1}^{T}  \sum_{n = 0}^N w_n\left[\boldsymbol{\psi}(t_{m}) - Q \exp(\Sigma t_m) Q^\dagger \boldsymbol{\psi}(t_0) \right]_n^2 \label{SIeq:optloss2}.
\end{equation}
Writing the formula in this way enables us to optimize or constrain the eigenvectors and eigenvalues of $A$ separately. To optimize $Q$ we follow the procedure in~\cite{sorber2015structured}, parameterizing the orthogonal matrix as the product of Householder matrices, $Q = H_1 H_2 \cdots H_N$, where each Householder matrix has the from 
\begin{equation}
    H_n = \begin{bmatrix} \mathbb{I}_{n - 1} & 0 \\ 0 & \mathbb{I}_{N - n + 1} - \frac{\mathbf{p}_n^\dagger \mathbf{p}_n }{\lvert \mathbf{p}_n\rvert^2}
\end{bmatrix} 
\label{eq:house_vec}
\end{equation}
and can be further parameterized by a vector $\mathbf{p}_n$ of length $N - n + 1$. Fast in-place matrix vector multiplication algorithms exist for both Householder matrices and $\exp(\Sigma t)$ enabling us to efficiently compute the loss function from the combined parameter vector $\mathbf{p} = [\mathbf{p}_1, \cdots \mathbf{p}_N, \lambda_1, \cdots, \lambda_r]$. To minimize the loss function we calculate gradients using automatic differentiation and perform gradient descent using both the AdaBelief algorithm followed by BFGS.

\subsection{Householder parameterization of the complex Steifel manifold (CSM)}
To parameterize the full model by the Hermitian Hamiltonian $H$, we decompose $H=U\Lambda U^\dagger$, where $U$ is unitary and $\Lambda$ is a diagonal matrix of real eigenvalues. We parameterize $U$ using Eq.~\eqref{eq:house_vec}, where now $\mathbf{p}_n$ is a complex vector. To learn the time dependent dynamics, each entry of $\mathbf{p}_n$ is parameterized by coefficients of a basis function expansion in time. $\Lambda$ is parameterized similarly, but here each entry is real.

\subsection{Model inference: practice}\label{SI:model_practice} 
The input data are the $(x,y)$ positions along the centerline of an object undergoing undulatory motion. We begin by performing a Chebyshev transformation at each time step to transform the real space data as a function of time and arc length to mode data as a function of time, $\hat{\boldsymbol{\psi}}$. Next, we calculate the rescaled mode vectors $\boldsymbol{\psi}$ from the scaled mode vectors $\hat{\boldsymbol{\psi}}$. 

If we focus on learning dynamics across a timescale of one oscillation of the propulsive body wave, the next step is to determine the length of one oscillation from the starting time point. We first compute the FFT for a range of data starting at the specified time and find the maximum frequency amplitude. We then search near the period corresponding to the maximum frequency to locate the most similar shape to the initial shape, computed by finding the L2 norm of the difference between the shape modes. The minimum difference is then considered to be the end of the oscillation. We set the eigenvalues of the Hamiltonian that we will infer to be integer multiples of this oscillation frequency.

The model inference is performed on Equations~\eqref{e:dynamics}. 
For straight motion, we first optimize the Hamiltonian with the eigenvalues constrained using the AdaBelief gradient-based optimization algorithm with forward-mode automatic differentation for gradient calculations. 
After this process, the BFGS algorithm is used to further optimize the Hamiltonian. 
Specifically, we are optimizing for $Q$ by parameterizing $Q$ with Householder matrices. 
For inferring time-depending Hamiltonians, we parameterize the Hamiltonian through its spectral decomposition $H=U\Lambda U^\dagger$, where $U$ is a unitary matrix and $\Lambda$ is a real diagonal matrix. $U$ is parameterized with complex Householder matrices while $\Lambda$ is parameterized through its individual diagonal elements. Each entry of the Householder vectors and $\Lambda$ diagonal entries are represented in time with a basis function expansion. In this work, we use a Chebyshev representation.
To fit time-dependent Hamiltonians for long time periods, the trajectory is split into individual 10 second trajectories.
The Hamiltonian is initialized by linearly interpolating between Hamiltonians fit to one period long straight trajectories beginning at each frame (0.05 seconds) of the trajectory. 
The Hamiltonian is then optimized using the same procedure as for straight motion, but now the eigenvalue constraint is lifted and the ODE is time-dependent. 
This process is repeated for the next 10 second long trajectory, where the worm shape initial condition for the next trajectory is set to be the final worm shape from the previous optimization.
After fitting the dynamics for the shape modes, the center of mass dynamics, described by $\boldsymbol{h}_0$, is optimized with the same optimization procedure used for the Hamiltonian.

\subsection{Loss function weighting}\label{SI:loss_weighting}
\begin{figure}[b]
    \centering
    \includegraphics[width=8cm]{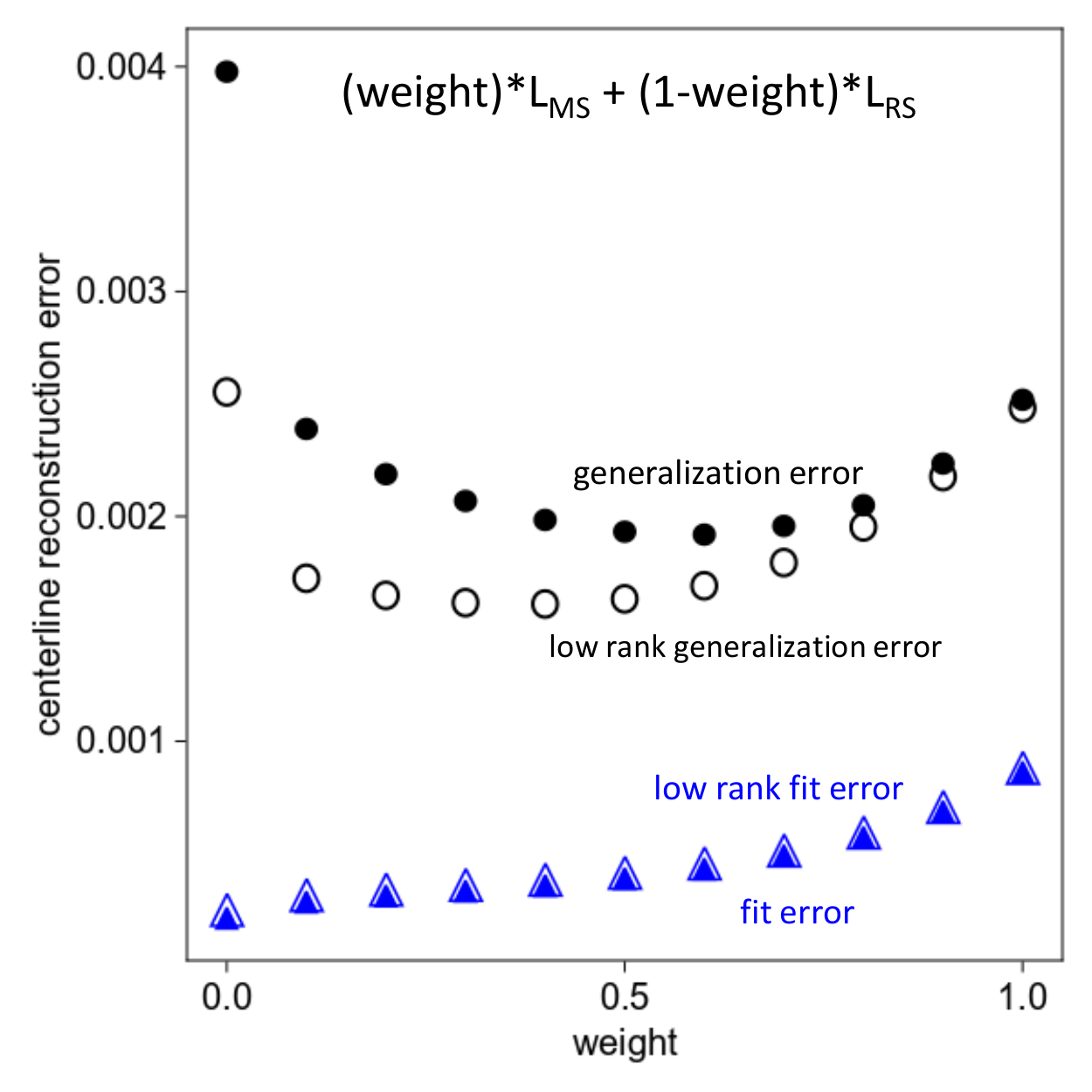}
    \caption{Centerline reconstruction error for models trained with different weightings of the real space and mode space loss in the combined loss function. Generalization error is the centerline reconstruction error after simulating the model on an initial condition different from the initial condition used in fitting. Fit error is the centerline reconstruction error when simulating the model on the initial condition from fitting.}
    \label{SIfig:error_fig}
\end{figure}
As described in the paper, we utilize a loss function that consists of a combined mode space and real space loss. The real space loss $L_{\text{RS}}(\mathbf{p}_H)$ calculates the mean square deviation between the CCOM subtracted field data $(x(s, t) - \hat{x}_0(t), y(s, t) - \hat{y}_0(t))$ and their prediction reconstructed from $\boldsymbol{\psi}_P$ calculated by integrating~\eqref{eq:SHAPEdynamics} with $H_1(\mathbf{p}_{H_1})$. The mode space loss $L_{\text{MS}}(\mathbf{p}_H)$ calculates the mean square deviation between the $\boldsymbol{\psi}_D$ calculated from the data and $\boldsymbol{\psi}_P$ normalized by the maximum standard deviations of all modes $\boldsymbol{\psi}_D$. The effect of the relative weighting of these losses is shown in Figure \ref{SIfig:error_fig}. As we increase the real space loss (move to the left on the x-axis), the centerline reconstruction error decreases (blue triangles). However, if we only have real space loss, the generalization error is very large (black circle in the top left corner), and the generalization error takes a minumum value for similar weighting of the two losses. Therefore, increasing the mode space loss promotes smaller generalization errors. In addition, while the low rank fit error is slightly larger than the full rank fit error (open versus closed triangles), the low rank generalization error is smaller than the full rank generalization error (open versus closed circles). These trends reveal that the mode space promotes generalizability while the real space loss promotes a good fit and the low rank model generalizes better than the full rank model. In the paper, we use a weight $=0.5$, which is between the minimum generalization error weights for the full and low rank models.

\section{Analytical solution to the model}\label{SI:analytical_solution}
We can use the eigendecomposition of the final $H_1$ matrix to derive the analytical solution for $\boldsymbol{\psi}$ and $\mathbf{r}(s,t) - \mathbf{r}_0(t)$, where $\mathbf{r}_0 = \boldsymbol{\psi}_0 T_0(s) = \boldsymbol{\psi}_0$ represents the dynamics coming from the constant Chebyshev polynomial $T_0 = 1$ (CCOM). 
For a $(2K + 1) \times (2K + 1)$ dimensional $H_1$ matrix we will have $1$ zero eigenvalue and a corresponding real eigenvector $\boldsymbol{\phi}_0$ and $K$ opposite sign pairs of real eigenvalues and their corresponding complex conjugate eigenvectors, $\lambda_1^\pm\,,\lambda_2^\pm\,,\cdots,\lambda_K^\pm$ and $\boldsymbol{\phi}_1^\pm\,,\boldsymbol{\phi}_2^\pm,\cdots,\boldsymbol{\phi}^\pm_K$. 
The complex eigenvector can be decomposed into a real and imaginary part, $\boldsymbol{\phi}_k=\boldsymbol{v}_k +i\boldsymbol{w}_k$.

Using this formulation, the analytical solution corresponding to a single eigenvalue and eigenvector pair can be written as 
\begin{equation*}
    \boldsymbol{s}_k^\pm = e^{i\lambda_k^\pm t}\boldsymbol{\phi}_k^\pm, \quad \boldsymbol{s}_0 = \boldsymbol{\phi}_0 = \boldsymbol{v}_0.
\end{equation*}
Using Euler's formula we get,
\begin{equation*}
    \boldsymbol{s}_k^\pm = (\cos{\lambda_k^\pm t}+i\sin{\lambda_k^\pm t})\boldsymbol{\phi}_k^\pm =  (\cos{\lambda_k^+t}\pm i\sin{\lambda_k^+t})(\boldsymbol{v}_k\pm i\boldsymbol{w}_k),
\end{equation*}
which separates into real and imaginary parts
\begin{equation*}
    \boldsymbol{s}_k^\pm =  (\cos{\lambda_k^+t}\boldsymbol{v}_k-\sin{\lambda_k^+t}\boldsymbol{w}_k) \pm i(\cos{\lambda_k^+t}\boldsymbol{w}_k+\sin{\lambda_k^+t}\boldsymbol{v}_k).
\end{equation*}
Defining $\tilde{\boldsymbol{v}}_k(t) = (\cos{\lambda_kt}\boldsymbol{v}_k-\sin{\lambda_kt}\boldsymbol{w}_k)$ and $\tilde{\boldsymbol{w}}_k(t) = (\cos{\lambda_kt}\boldsymbol{w}_k+\sin{\lambda_kt}\boldsymbol{v}_k)$ gives
\begin{equation*}
    \boldsymbol{s}_k^\pm = \tilde{\boldsymbol{v}}_k(t) \pm i\tilde{\boldsymbol{w}}_k(t).
\end{equation*}
Each pair of $\tilde{\boldsymbol{v}}_k(t)$ and $\tilde{\boldsymbol{w}}_k(t)$ are two linearly independent solutions so we can write the general solution as 
\begin{equation*}
    \boldsymbol{s} = c_{0, 1} \boldsymbol{v}_0 + \sum_{k=1}^K c_{k,1}\tilde{\boldsymbol{v}}_k(t) + c_{k,2}\tilde{\boldsymbol{w}}_k(t).
\end{equation*}
Since both $\boldsymbol{\psi}_x$ and $\boldsymbol{\psi}_y$ satisfy the same dynamical equation we get the general solution for $\boldsymbol{\psi}(t)$,
\begin{equation} \label{SIeq:general_ms_solution}
    \boldsymbol{\psi}(t) = \begin{bmatrix}
    \boldsymbol{\psi}_x\\
    \boldsymbol{\psi}_y
    \end{bmatrix}
    = 
    \begin{bmatrix}
    c_{0,1}\boldsymbol{v}_0(t) \\
    d_{0,1}\boldsymbol{v}_0(t)
    \end{bmatrix} + \sum_{k=1}^K \begin{bmatrix}
    c_{k,1}\tilde{\boldsymbol{v}}_k(t) + c_{k,2}\tilde{\boldsymbol{w}}_k(t)\\
    d_{k,1}\tilde{\boldsymbol{v}}_k(t) + d_{k,2}\tilde{\boldsymbol{w}}_k(t)
    \end{bmatrix}.
\end{equation}
We now convert Eq.~\eqref{SIeq:general_ms_solution} to real space. First, we undo the $L$ scaling to go from the hypersphere back to the hyperellipsoid, 
\begin{equation}
\hat{\boldsymbol{\psi}}(t) = \begin{bmatrix}
    \hat{\boldsymbol{\psi}}_x\\
    \hat{\boldsymbol{\psi}}_y
    \end{bmatrix}
    = 
    \ell \begin{bmatrix}
    c_{0,1}(L^{-1})^\dagger \boldsymbol{v}_0(t) \\
    d_{0,1}(L^{-1})^\dagger \boldsymbol{v}_0(t)
    \end{bmatrix} + \ell\sum_{k=1}^K \begin{bmatrix}
    c_{k,1}(L^{-1})^\dagger \tilde{\boldsymbol{v}}_k(t) + c_{k,2}(L^{-1})^\dagger \tilde{\boldsymbol{w}}_k(t)\\
    d_{k,1}(L^{-1})^\dagger \tilde{\boldsymbol{v}}_k(t) + d_{k,2}(L^{-1})^\dagger \tilde{\boldsymbol{w}}_k(t)
    \end{bmatrix} \label{SIeq:psi_hat_solution}
\end{equation}
then the real space solution can be written for $x(s, t)$ as, 
\begin{equation*}
    x(s, t) = \sum_{n = 0}^N \hat{x}_n(t) T_n(s) = \hat{x}_0(t) + \mathbf{T}(s)^\dagger \hat{\boldsymbol{\psi}}_x(t)
\end{equation*}
where we use the fact that $T_0(s) = 1$ and define the Chebyshev vector $\mathbf{T}(s) = [T_1(s),\, T_2(s),\, \cdots,\,T_N(s)]$. Substituting in the solution~\eqref{SIeq:psi_hat_solution} we get, 
\begin{subequations}
\begin{align}
x(s, t) - \hat{x}_0(t) &= \ell \boldsymbol{T}(s)^\dagger (L^{-1})^\dagger \left[ c_{0, 1}  \boldsymbol{v}_0 + \sum_{k = 1}^K c_{k, 1} \tilde{\boldsymbol{v}}_k + c_{k, 2} \tilde{\boldsymbol{w}}_k \right] \nonumber
\intertext{using the definition of $\tilde{\boldsymbol{v}}_k$ and $\tilde{\boldsymbol{w}}_k$ the equation becomes,}
&= \ell \left[L^{-1}\mathbf{T}(s)\right]^\dagger \left[c_{0, 1} \boldsymbol{v}_0 + \sum_{k = 1}^K c_{k, 1} (\cos{(\lambda_k^+t)}\boldsymbol{v}_{k}-\sin{(\lambda_k^+t)}\boldsymbol{w}_{k}) + c_{k, 2} (\cos{(\lambda_k^+t)}\boldsymbol{w}_{k}+\sin{(\lambda_k^+t)}\boldsymbol{v}_{k}) \right] \nonumber \\
x(s, t) - \hat{x}_0(t) &= \ell \left[L^{-1}\mathbf{T}(s)\right]^\dagger \left[c_{0, 1} \boldsymbol{v}_0 + \sum_{k = 1}^K \cos{(\lambda_k^+t)}(c_{k, 1}\boldsymbol{v}_{k} + c_{k, 2} \boldsymbol{w}_{k}) +  \sin{(\lambda_k^+t)} (c_{k, 2}\boldsymbol{v}_{k} - c_{k, 1} \boldsymbol{w}_k) \right] \nonumber, 
\intertext{finally we can write this in terms of the real space eigenfunctions $v_k(s) = \mathbf{T}^\dagger(s) (L^{-1})^\dagger \boldsymbol{v}_k$ and $w_k(s) = \mathbf{T}^\dagger(s) (L^{-1})^\dagger \boldsymbol{w}_k$}
x(s, t) - \hat{x}_0(t) &= \ell \left[c_{0, 1} v_0(s) + \sum_{k = 1}^K \cos{(\lambda_k^+t)}(c_{k, 1}v_{k}(s) + c_{k, 2} w_{k}(s)) +  \sin{(\lambda_k^+t)} (c_{k, 2}v_{k}(s) - c_{k, 1} w_k(s)) \right],
\intertext{a similar calculation for $y(s, t)$ yields, }
y(s, t) - \hat{y}_0(t) &= \ell \left[d_{0, 1} v_0(s) + \sum_{k = 1}^K \cos{(\lambda_kt)}(d_{k, 1}v_{k}(s) + d_{k, 2} w_{k}(s)) +  \sin{(\lambda_kt)} (d_{k, 2}v_{k}(s) - d_{k, 1} w_k(s)) \right].
\end{align}
\end{subequations}

Now consider the state where we enforce that the $\lambda_k = k \lambda$ are integer multiples of some base frequency $\lambda$
\begin{equation}
    x(s, t) - \hat{x}_0(t) = \ell \left[c_{0, 1} v_0(s) + \sum_{k = 1}^K \cos{(k\lambda t)}(c_{k, 1}v_{k}(s) + c_{k, 2} w_{k}(s)) +  \sin{(k \lambda t)} (c_{k, 2}v_{k}(s) - c_{k, 1} w_k(s)) \right]. \label{SIeq:x_integerlambda_generalsol}
\end{equation}
We define the time-average square deviation from $\hat{x}_0(t)$ as
\begin{equation*}
\left\langle \left(x(s, t) - \hat{x}_0(t)\right)^2 \right\rangle_t(s) = \frac{\lambda}{2\pi} \int_{0}^{\frac{2\pi}{\lambda}} \mathrm{d}t\,\left(x(s, t) - \hat{x}_0(t)\right)^2.
\end{equation*}
Noting the orthogonality of $\sin$ and $\cos$ means that the only trigonometric functions that have non-zero time average are of the form $\cos(l \lambda t)^2$ and $\sin(l \lambda t)^2$, the can substitute~\eqref{SIeq:x_integerlambda_generalsol} into the definition for the squared deviation and only keep non-zero terms,
\begin{subequations}
\begin{align}
    \left\langle \left(x(s, t) - \hat{x}_0(t)\right)^2 \right\rangle_t(s) &= \frac{\ell^2 \lambda}{2\pi} \int_{0}^{\frac{2\pi}{\lambda}} \mathrm{d}t\, c_{0, 1}^2 v_0(s)^2 + \sum_{k = 1}^K  \left[(c_{k, 1} v_{k}(s) + c_{k, 2} w_{k}(s)) \right]^2 \cos(k \lambda t)^2 \nonumber
    \\ & \hspace{8em} + \sum_{k = 1}^K \left[c_{k, 2}v_{k}(s) - c_{k, 1} w_{k}(s) \right]^2 \sin(k \lambda t)^2 \nonumber \\
    &= \ell^2 c_{0, 1}^2 v_0(s)^2  + \frac{\ell^2}{2}\sum_{k = 1}^K \left(c_{k, 1}^2 + c_{k, 2}^2\right) \left(v_k(s)^2 + w_k(s)^2\right).
\end{align}
Similarly we have for $y(s, t)$ we have, 
\begin{align}
    \left\langle \left(y(s, t) - \hat{y}_0(t)\right)^2 \right\rangle_t(s) = \ell^2 d_{0, 1}^2 v_0(s)^2  + \frac{\ell^2}{2}\sum_{k = 1}^K \left(d_{k, 1}^2 + d_{k, 2}^2\right) \left(v_k(s)^2 + w_k(s)^2\right).
\end{align}
This means that the contribution to the mean square deviation corresponding to each eigenvalue takes the form of a spatial density $\rho_k(s) = v_k(s)^2 + w_k(s)^2$ with corresponding weight $c_{0, 1}^2 + d_{0, 1}^2$ for $k = 0$ and $c_{k, 1}^2 + c_{k, 2}^2 + d_{k, 1}^2 + d_{k, 2}^2$ for $k \ge 1$.

We can readily extend this time average to a square deviation from some time-constant line $l(s)$, 
\begin{align}
    \left\langle \left(x(s, t) - \hat{x}_0(t) - l(s)\right)^2 \right\rangle_t(s) &= \left(\ell c_{0, 1} v_0(s) - l(s)\right)^2  + \frac{\ell^2}{2}\sum_{k = 1}^K \left(c_{k, 1}^2 + c_{k, 2}^2\right) \left(v_k(s)^2 + w_k(s)^2\right).
\end{align}
\end{subequations}
For example, if $l(s) = \ell c_{0, 1} v(s)$, an approximation to the center line of the worm, the first term vanishes meaning the dominant term in the deviation is, 
\begin{equation*}
\left\langle \left(x(s, t) - \hat{x}_0(t) - \ell c_{0, 1} v_0(s) \right)^2 \right\rangle_t(s) \approx \frac{\ell^2}{2}\left(c_{1, 1}^2 + c_{1, 2}^2\right) \left(v_1(s)^2 + w_1(s)^2\right)
\end{equation*}

Note since $\boldsymbol{v}_k$ and $\boldsymbol{w}_k$ are the columns or our orthogonal matrix $Q$ in the spectral decomposition $\boldsymbol{v}_k^\dagger \boldsymbol{v}_l = \delta_{l, k}$, $\boldsymbol{w}_k^\dagger \boldsymbol{w}_l = \delta_{l, k}$ and $\boldsymbol{v}_k^\dagger \boldsymbol{w}_l = 0$. This implies that, \begin{align}
    \int_{-1}^{1} w(s) v_l(s) v_k(s) &= \sum_{n = 1}^N \sum_{m = 1}^N \left[(L^{-1})^\dagger \boldsymbol{v}_k\right]_n \left[(L^{-1})^\dagger \boldsymbol{v}_l\right]_m \int_{-1}^{1}  \mathrm{d}s \, w(s) T_n(s) T_m(s)  
    \notag\\ 
    &= \frac{\pi}{2}\sum_{n = 1}^N \left[(L^{-1})^\dagger \boldsymbol{v}_k\right]_n \left[(L^{-1})^\dagger \boldsymbol{v}_l\right]_n
\end{align}

\section{Analytical solution to the turning model}\label{SI:analytical_solution_full}

We consider the full dynamics, 
\begin{equation}
\dot{\boldsymbol{\psi}} = \begin{bmatrix} A & S \\ - S & A \end{bmatrix} \boldsymbol{\psi} .
\end{equation}
We need the eigendecomposition of the full matrix. Assume that $\mathbf{w} = [\mathbf{v}, -i\mathbf{v}]$ is an eigenvector. Then we get the following set of conditions,
\begin{equation}
\begin{bmatrix} A & S \\ - S & A \end{bmatrix} \begin{bmatrix} \mathbf{v} \\ -i\mathbf{v} \end{bmatrix} = \begin{bmatrix} (A -  iS) \mathbf{v} \\ (-S - iA)\mathbf{v}\end{bmatrix} = \lambda \begin{bmatrix} \mathbf{v} \\ -i\mathbf{v} \end{bmatrix}
\end{equation}
requiring $(A - iS)\mathbf{v} = \lambda \mathbf{v}$. We see that $\mathbf{w}$ is an eigenvector of the larger matrix and the subvectors are the eigenvectors of $(A - iS)$. Since $A - iS$ is skew-Hermitian it is diagonalizable by a unitary matrix and has purely imaginary eigenvalues. Therefore, we can produce $n$ distinct complex eigenvectors of the form $\mathbf{w}_k = [\mathbf{v}_k, -i\mathbf{v}_k]$ and the vectors $\mathbf{v}_k$ can be stacked into a unitary matrix $V$, such that $VV^\dagger = V^\dagger V = I$. We can produce an additional $n$ eigenvectors since the matrix is real and, as a result, the complex conjugates of the eigenvectors must also be eigenvectors with complex conjugate eigenvalues, yielding a full set of $2n$ distinct eigenvectors:
\begin{equation*}
\mathbf{w}_k = \begin{bmatrix} \mathbf{v}_k\\ -i\mathbf{v}_k\end{bmatrix} \qquad \text{and} \qquad \mathbf{w}_k^* = \begin{bmatrix} \mathbf{v}_k^*\\ i\mathbf{v}_k^*\end{bmatrix} 
\end{equation*}
Note this is the same matrix as in the complex formulation. 
We can write this in block matrix form as follows,
\begin{equation} \label{SIeq:full_eigdecomp}
M = \frac{1}{2}\begin{bmatrix} V & V^* \\ -iV & iV^* \end{bmatrix} \begin{bmatrix} 
i\Lambda & 0 \\ 0 & -i\Lambda\end{bmatrix} \begin{bmatrix} (V^*)^\top & i(V^*)^\top \\ V^\top & -iV^\top \end{bmatrix} 
\end{equation}
where $V$ is a unitary matrix above and $\Lambda$ is a real diagonal matrix and we have made use of the fact that, 
\begin{equation*}
W = \begin{bmatrix} V & V^* \\ -iV & iV^* \end{bmatrix} 
\end{equation*}
is also unitary up to a scaling factor of $2$,
\begin{align*}
WW^\dagger &= \begin{bmatrix} V & V^* \\ -iV & iV^* \end{bmatrix} \begin{bmatrix} V^\dagger & iV^\dagger \\ (V^*)^\dagger & -i(V^*)^\dagger \end{bmatrix} = \begin{bmatrix} VV^\dagger + (VV^\dagger)^* & iVV^\dagger - i (VV^\dagger)^* \\ -iVV^\dagger + i (VV^\dagger)^* & VV^\dagger + (VV^\dagger)^*\end{bmatrix} = 2 \begin{bmatrix} \mathbb{I}_n & 0 \\ 0 & \mathbb{I}_n\end{bmatrix} = 2\mathbb{I}_{2n} \\
W^\dagger W &=  \begin{bmatrix} V^\dagger & iV^\dagger \\ (V^*)^\dagger & -i(V^*)^\dagger \end{bmatrix} \begin{bmatrix} V & V^* \\ -iV & iV^* \end{bmatrix} = \begin{bmatrix} 2V^\dagger V & V^\dagger V^* -  V^\dagger V^* \\ (V^*)^\dagger V -  (V^*)^\dagger V & 2(V^\dagger V)^* \end{bmatrix} = 2 \begin{bmatrix} \mathbb{I}_n & 0 \\ 0 & \mathbb{I}_n\end{bmatrix} = 2\mathbb{I}_{2n}.
\end{align*}
We can rewrite this only in terms of real matrices using the identity, 
\begin{equation*}
\begin{bmatrix} i\Lambda & 0 \\ 0 & -i\Lambda \end{bmatrix} = \frac{1}{2} \begin{bmatrix} \mathbb{I}_n & -i\mathbb{I}_n \\ \mathbb{I}_n & i\mathbb{I}_n \end{bmatrix} \begin{bmatrix} 0 & \Lambda \\ -\Lambda & 0\end{bmatrix} \begin{bmatrix} \mathbb{I}_n & \mathbb{I}_n \\ i\mathbb{I}_n & -i\mathbb{I}_n \end{bmatrix}.
\end{equation*}
The eigendecomposition then becomes, 
\begin{align*}
M &= \frac{1}{4}\begin{bmatrix} V & V^* \\ -iV & iV^* \end{bmatrix} \begin{bmatrix} \mathbb{I}_n & -i\mathbb{I}_n \\ \mathbb{I}_n & i\mathbb{I}_n \end{bmatrix} \begin{bmatrix} 0 & \Lambda \\ -\Lambda & 0\end{bmatrix} \begin{bmatrix} \mathbb{I}_n & \mathbb{I}_n \\ i\mathbb{I}_n & -i\mathbb{I}_n \end{bmatrix} \begin{bmatrix} (V^*)^\top & i(V^*)^\top \\ V^\top & -iV^\top \end{bmatrix}  \\
&= \frac{1}{4} \begin{bmatrix} V + V^* & -i (V -V^*) \\ -i (V -V^*) & -(V + V^*) \end{bmatrix}  \begin{bmatrix} 0 & \Lambda \\ -\Lambda & 0\end{bmatrix} \begin{bmatrix} (V + V^*)^\top & -i (V -V^*)^\top \\ -i (V -V^*)^\top & -(V + V^*)^\top \end{bmatrix} \\
&= \begin{bmatrix} X & Y \\ Y & -X \end{bmatrix} \begin{bmatrix} 0 & \Lambda \\ -\Lambda & 0\end{bmatrix} \begin{bmatrix} X^\top & Y^\top \\ Y^\top & -X^\top \end{bmatrix}  = Z \begin{bmatrix} 0 & \Lambda \\ -\Lambda & 0 \end{bmatrix} Z^\top
\end{align*}
where we write $V = X + iY$. Note that since $VV^\dagger = V^\dagger V = \mathbb{I}_n$ we get the following conditions on $X$ and $Y$, $VV^\dagger = (X + iY)(X^\top - i Y^\top) = XX^\top + YY^\top + i (YX^\top -XY^\top) = \mathbb{I}_n$ which implies that $XX^\top + YY^\top = \mathbb{I}_n$ and $YX^\top = XY^\top$. $V^\dagger V = \mathbb{I}_n$ gives the additional constraints  $X^\top X + Y^\top Y = \mathbb{I}_n$ and $X^\top Y = Y^\top X$. From these conditions we can easily verify that the matrix $Z$ is orthogonal. 

We can use the eigendecomposition to find the analytic solution of our ODE problem. The orthogonality of $Z$ means that we can write the matrix exponential of $M$ as, 
\begin{equation*} \exp(tM) = Z \exp\left(t \begin{bmatrix} 0 & \Lambda \\ -\Lambda & 0\end{bmatrix} \right) Z^\top = Z \begin{bmatrix} \cos(t\Lambda) & \sin(t\Lambda)  \\ -\sin(t\Lambda)  & \cos(t\Lambda) \end{bmatrix} Z^\top\end{equation*} 
We can permute the order of the eigenvectors by multiplying by a permutation matrix. Doing so we can rewrite the matrix and its exponential in the following form, 
\begin{equation*}
M = 
    \begin{bmatrix} Z_1 & Z_2 & \cdots & Z_n \end{bmatrix}
    \begin{bmatrix} \Lambda_1 & & & \\ & \Lambda_2 && \\ && \ddots & \\ &&& \Lambda_n \end{bmatrix}
    \begin{bmatrix} Z_1^\top \\ Z_2^\top \\ \vdots \\ Z_n^\top \end{bmatrix}
\end{equation*}
and
\begin{align*}
    \exp(tM) &= 
    \begin{bmatrix} Z_1 & Z_2 & \cdots & Z_n \end{bmatrix}
    \begin{bmatrix} \exp(t \Lambda_1) & & & \\ & \exp(t \Lambda_2) && \\ && \ddots & \\ &&& \exp(t \Lambda_n) \end{bmatrix}
    \begin{bmatrix} Z_1^\top \\ Z_2^\top \\ \vdots \\ Z_n^\top \end{bmatrix}  
    = \sum_{k = 1}^n Z_k \exp(t \Lambda_k) Z_k^\top
\end{align*}
where we introduce the following $2n \times 2$ matrix, 
\begin{equation*}
    Z_k = \begin{bmatrix}
    \mathbf{x}_k & \mathbf{y}_k \\ \mathbf{y}_k & -\mathbf{x}_k
    \end{bmatrix} = \begin{bmatrix} \mathbf{z}_k & \mathbb{P}_n \mathbf{z}_k \end{bmatrix}, \text{ where } \mathbb{P}_n = \begin{bmatrix} 0 & \mathbb{I}_n \\ -\mathbb{I}_n & 0 \end{bmatrix}
\end{equation*}
and $\mathbf{z}_k = [\mathbf{x}_k, \mathbf{y}_k]$. $\mathbb{P}_n$ has the following properties, $\mathbb{P}_n^\top = -\mathbb{P}_n$ and $\mathbb{P}_n^2 = -I_{2n}$. The conditions on $X$ and $Y$ mean that $Z_l^\top Z_k = \delta_{l,k}\mathbb{I}_2$.
The solution of our dynamical sytstem is then given by,
\begin{equation}
\boldsymbol{\psi}(t) =  \exp(t M) \boldsymbol{\psi}_0 = \sum_{k = 1}^n Z_k \exp(t \Lambda_k) (Z_k^\top \boldsymbol{\psi}_0).
\end{equation}

\section{Straight and Turning Components of the Hamiltonian}

\begin{figure}
    \centering
    \includegraphics[width=0.5\textwidth]{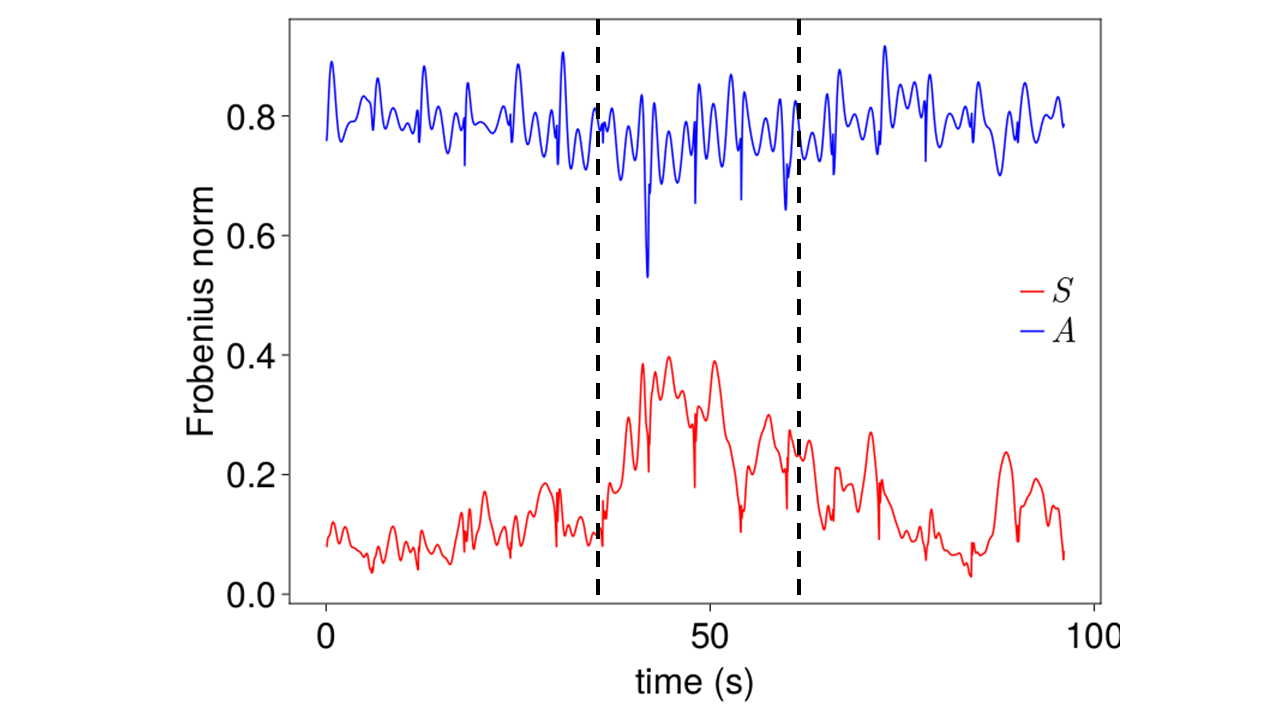}
    \caption{Frobenius norm of the straight ($A$) and turning ($S$) components of the Hamiltonian $H$. The norm of turning component increases during a turn.}
    \label{SIfig:AS_norm1}
\end{figure}

As discussed in the main text, $S$ represents the turning component of the Hamiltonian, while $A$ represents the straight component of the Hamiltonian. This is demonstrated in Fig.~S7, which shows the Frobenius norm of $S$ and $A$ during a worm trajectory, where a reversal and turn section is indicated with dotted lines.

\section{Grassmann distance}\label{SI:grassmann}
To calculate the distances between the subspaces spanned by the real and imaginary parts of the eigenvector corresponding to the smallest nonzero eigenvalues~(Fig.~\ref{fig:fig4}), we use the Grassmann distance. The Grassmann distance between two subspaces can be calculated by
\begin{equation}
    d_G(A,B)=\sqrt{\sum_i \theta_i^2},
\end{equation}
where $A$ and $B$ are two matrices whose columns are an orthonormal basis of their respective subspaces and $\theta_i$ are the principal angles between $A$ and $B$~\cite{ye2016schubert}. The principal angles can be calculated through an SVD, where the singular values of $A^\top B$ are $\sigma_i = \cos{(\theta_i)}$. 

\end{document}